\documentclass[aps,pre,superscriptaddress,twocolumn,floatfix]{revtex4-1}
\usepackage{graphicx}
\usepackage{amsmath}
\usepackage{amssymb}
\usepackage{bm}
\usepackage{bbm}
\usepackage[normalem]{ulem}
\usepackage{dcolumn}
\usepackage{color}
\usepackage{xcolor}
\usepackage{xfrac}
\usepackage{mathtools}
\usepackage{hyperref}

\begin{document}

\title{Structure of Quantum Mean Force Gibbs States for Coupled Harmonic Systems }

\author{Joonhyun Yeo} 
\affiliation{Department of Physics, Konkuk University, Seoul 05029, Korea}
\author{Haena Shim}
\affiliation{Department of Physics, Konkuk University, Seoul 05029, Korea}

\date{\today}

\begin{abstract}
An open quantum system interacting with a heat bath at given temperature is expected to reach the
mean force Gibbs (MFG) state as a steady state. The MFG state is given by tracing out the bath degrees of freedom from the 
equilibrium Gibbs state of the total system plus bath. When the interaction between 
the system and the bath is not negligible, it is different from the usual system Gibbs state obtained from the system Hamiltonian only.
Using the path integral method, we present the exact MFG state for a coupled system of quantum harmonic oscillators
in contact with multiple thermal baths at the same temperature. We develop a nonperturbative
method to calculate the covariances with respect to the MFG state. By comparing them with those obtained 
from the system Gibbs state, we find that the effect of coupling
to the bath decays exponentially as a function of the distance from the system-bath boundary.
This is similar to the skin effect found recently for a quantum spin chain interacting with an environment. 
Using the exact results, we also investigate the ultrastrong coupling limit where the coupling between the system and the bath gets
arbitrarily large and make a connection with the recent result found for a general quantum system.

\end{abstract}



\maketitle

\section{Introduction}

In the standard theory of thermodynamics the interaction between a system and its environment
is assumed to be weak. However, for systems with decreasing sizes \cite{Jarzynski_2017} or for quantum 
mechanical systems \cite{Ford_1985}, the
interaction cannot be neglected. This consideration has led to the recent development of 
strong coupling thermodynamics 
\cite{Campisi_2009, Gelin_2009, subasi_equilibrium_2012, seifert_2016, Jarzynski_2017, Strasberg_Esposito_2017, Aurell_2017, miller2018, Hsiang_Hu_2018, Perarnau_2018, Talkner_2020, rivas_strong_2020, Anto_2023, Kaneyasu_2023, Diba_2024}. 
The mean force Gibbs (MFG) state \cite{trushechkin_open_2022, cresser_weak_2021,Chiu_2022}
and the associated Hamiltonian of mean force \cite{Kirkwood_1935,Grabert_Weiss_Talkner_1984, hilt_hamiltonian_2011,miller2018,timofeev2022hamiltonian} play a central role in the theory of strong coupling 
thermodynamics.
The MFG state describes the equilibrium state of the system interacting with a thermal bath at given temperature. 
When the interaction between the system and the bath is not negligible, the MFG state deviates from the usual system 
Gibbs state which is described by the system Hamiltonian only \cite{trushechkin_open_2022}, and the Hamiltonian of mean force is different
from the system Hamiltonian \cite{hilt_hamiltonian_2011}.

Despite the importance in strong coupling thermodynamics, explicit evaluation of the quantum MFG state turns out
to be quite difficult. Exact results are known only for simple limited cases \cite{Grabert_Weiss_Talkner_1984, hilt_hamiltonian_2011}
of a damped harmonic oscillator described by the Caldeira-Leggett model \cite{Caldeira_1983}. Recently, the
MFG state for a general quantum system interacting with a bosonic environment
was obtained in both weak and ultrastrong coupling limits \cite{cresser_weak_2021}. 
The ultrastrong coupling limit of the MFG state obtained in Ref.~\cite{cresser_weak_2021} is quite interesting as
it is characterized by a diagonal projection of the system Hamiltonian onto the 
eigenstates of the system operator which couples to the environment. 
In order to gain further understanding of quantum thermodynamics beyond weak coupling, it would be useful to have 
explicit expressions of the MFG state of a more general quantum system which covers the range of the 
coupling strength from weak to ultrastrong. 
One of the purposes of this paper is to provide such examples of the quantum 
MFG state.

Another aspect of the MFG state which has not been investigated as yet is 
its spatial structure. 
For an open quantum system with spatial extent, only a part of the system
is directly coupled to the heat bath. Therefore, the MFG state as an equilibrium state
is expected to exhibit a spatial structure reflecting this local coupling. 
Recently, in Ref.~\cite{burke_structure_2024}, the Hamiltonian of mean force of a spin chain interacting with its environment
was studied numerically.
The results show a kind of skin effect where the effect of the coupling to the environment decays exponentially
as one moves away from the system-environment boundary. This suggests that the effect of the local coupling to the environment
on the equilibrium state of the system is confined to the vicinity of the point of contact with the environment.

In this paper, we present calculations leading to the MFG state of a coupled system of quantum mechanical harmonic oscillators
interacting with multiple thermal baths maintained at given temperature. This is an exact result which covers
the range of the coupling strength to the heat baths from
weak to ultrastrong. We consider an open quantum system of coupled oscillators described by
the Caldeira-Leggett type Hamiltonian \cite{Caldeira_1983} where only parts of the system are
coupled with the baths. By explicitly evaluating the path integrals involved in tracing out the bath degrees of freedom, we 
present a formalism of obtaining the exact MFG state. Being a Gaussian state, the MFG state is completely described
by its covariances. We present a nonperturbative method to evaluate the covariances with respect to the  
MFG state. By studying the difference in the covariances
obtained with respect to the MFG and the system Gibbs states, we show
how the local nature of the coupling to the bath shows up 
in the MFG state. 
We find that there is a skin effect similar to that found in Ref.~\cite{burke_structure_2024},
where the effect of the coupling to the heat bath decays exponentially as a function of the distance from
the system-bath boundary. Using the exact expressions, we also explore the ultrastrong coupling regime.
We find that the expression found in Ref.~\cite{cresser_weak_2021} for this limit is still valid even 
in the present case where the coupling system operator is local and has a continuous spectrum.

In the following section, 
we present a general formalism for the calculation of the MFG state of coupled harmonic systems using the path integral method.  
In Sec.~\ref{sec:single}, we focus on the case where the system is in contact with a single heat bath.
Using a specific example of a ring of coupled oscillators, we present detailed expressions for the exact MFG covariances
and study their ultrastrong coupling limit. 
We then generalize to the case where the system is in contact with two or more baths at the same temperature
in Sec.~\ref{sec:multiple}. 
Finally. we conclude with discussion.

\section{MFG state for coupled harmonic systems from path integral}
\label{sec:setup}

In this section, we present the path integral approach to calculate the MFG state of coupled harmonic systems
in contact with multiple thermal baths at the same temperature.
The path integral method has been used extensively for the study of many aspects of open quantum systems 
\cite{grabert_quantum_1988, makri_1995, ishizaki_quantum_2005, martinez_dynamics_2013, estrada_quantum, richter_2017, funo_2018, yeo_2019}. 
The main ingredient of this formalism is the calculation of the reduced density operator
in terms of the so-called Feynman-Vernon influence functional \cite{Feynman_1963} which is obtained 
after tracing out the bath degrees of freedom.

An open quantum system in contact with a thermal bath
is expected to approach in the long time limit its stationary state described by the MFG state \cite{subasi_equilibrium_2012}. 
We consider a quantum mechanical system described by the Hamiltonian $H_{\rm S}$
interacting with the bath given by $H_{\rm B}$. 
The total Hamiltonian is written as
\begin{align}
\hat{H}=\hat{H}_{\rm S}+\hat{H}_{\rm B}+\hat{H}_{\rm I},
\label{total}
\end{align}
where the interaction between the system and the bath is governed by $H_{\rm I}$. 
The MFG state is defined by
\begin{align}
	\hat{\rho}_{\rm mfG} = \mathrm{Tr_B} \hat{\rho}_\beta \equiv \frac{1}{Z_\beta}{\rm Tr}_{\rm B} (e^{-\beta \hat{H}}) , \label{eq:rG}
\end{align}
which is obtained by the partial trace of the equilibrium state of the total system $\hat{\rho}_\beta$
with $Z_\beta\equiv \mathrm{Tr_{SB}} \exp(-\beta H)$.
When the interaction between the system and bath is not negligible, the MFG state is in general different from the 
usual system Gibbs state
\begin{align}
\hat{\rho}_{\rm G}=\frac{1}{\mathrm{Tr_S} e^{-\beta \hat{H}_{\rm S}}}e^{-\beta \hat{H}_{\rm S}}
\end{align}
given by the system Hamiltonian only.

We consider a collection of $N$ quantum harmonic oscillators of equal mass $M$. The coupling among these
oscillators is described by a symmetric positive definite matrix $\bm{\Lambda}$. The system is described by the Hamiltonian
\begin{align}
\hat{H}_{\rm S}=\sum_{i=0}^{N-1}\frac{\hat{p}_i^2}{2M}+\frac 1 2 \sum_{i,j=0}^{N-1}\Lambda_{ij}\hat{x}_i\hat{x}_j,
\label{Hs}
\end{align}
where $\hat{x}_i$ and $\hat{p}_i$ are the position and momentum operators of the $i$-th oscillator, respectively.

We consider the case where only a part of the system 
is in contact with a collection of heat baths labelled by $\alpha$. 
We set up a Caldeira-Leggett model \cite{Caldeira_1983} where
each bath is composed of harmonic oscillators with mass $m_{n,\alpha}$ and frequency $\omega_{n,\alpha}$.
Here $n=1,2,\cdots$ labels a collection of a large number of the harmonic oscillators which constitutes each heat bath. The distribution of the
masses and frequencies of these oscillators will be specified below.
The Hamiltonian for the heat baths and their interaction with the system can be written as
\begin{align}
&\hat{H}_{\rm B}+\hat{H}_{\rm I}   \label{HBI} \\
=&\sum_{\alpha\in\mathcal{B}} \sum_n \left[  \frac{\hat{\pi}^2_{n,\alpha}}{2m_{n,\alpha}}+\frac 1 2 m_{n,\alpha} \omega^2_{n,\alpha} 
\left(\hat{q}_{n,\alpha} - \frac{\kappa_{n,\alpha}}{m_{n,\alpha} \omega^2_{n,\alpha}}\hat{x}_\alpha\right)^2\right] , \nonumber
\end{align}
where $\hat{\pi}_{n,\alpha}$ and $\hat{q}_{n,\alpha}$ are the momentum and position operators for the oscillators in the bath $\alpha$, respectively.
Here the index $\alpha$ runs through a subset $\mathcal{B}$ of $\{0,1,\cdots,N-1\}$ such that the oscillator described by the position $\hat{x}_\alpha$
is coupled with the oscillators in the $\alpha$-th bath with the coupling constant $\kappa_{n,\alpha}$.
The distribution of the $\alpha$-th bath modes is conveniently described by
the spectral function
\begin{align}
J_\alpha(\omega)=\sum_n \frac{\kappa^2_{n,\alpha}}{2m_{n,\alpha}\omega_{n,\alpha}}\delta(\omega-\omega_{n,\alpha}).
\label{Ja}
\end{align}

We first write the matrix element of $\rho_\beta$ using Euclidian path integrals as
\begin{align}
& \langle \bm{\bar{x}},\{\bm{\bar{q}}_\alpha\} \vert
 \hat{\rho}_\beta \vert \bm{\bar{x}}',\{\bm{\bar{q}}' _\alpha\}\rangle \nonumber \\
=& \frac 1{Z_\beta}\int_{\bm{x}(0)=\bm{\bar{x}}'}^{\bm{x}(\beta\hbar)=\bm{\bar{x}}}\mathcal{D}\bm{x}(\tau) 
 \prod_{\alpha\in\mathcal{B}} \int_{\bm{q}_\alpha(0)=\bm{\bar{q}}'_\alpha}^{\bm{q}_\alpha(\beta\hbar)=\bm{\bar{q}}_\alpha}\mathcal{D}\bm{q}_\alpha(\tau)\ \nonumber \\
 &\times \exp\left [-\frac 1 {\hbar}\left( S_{\rm S}[\bm{x}]+S_{\rm B}[\{\bm{q}_\alpha\}]+S_{\rm I}[\bm{x},\{\bm{q}_\alpha\}]\right)\right],
\end{align}
where the system Euclidian action is given by
 \begin{align}
 S_{\rm S}[\bm{x}]=\int_0^{\beta\hbar} d\tau &\; \Big[ \frac M 2 \frac{d \bm{x}^{\rm T}(\tau)} {d\tau}
\frac{d \bm{x}(\tau)} {d\tau} +\frac 1 2 \bm{x}^{\rm T}(\tau)\bm{\Lambda}\bm{x}(\tau)\Big]  \label{Ss_mul}
\end{align} 
with the matrix notation $\bm{x}^{\rm T}=(x_0,x_1,\cdots,x_{N-1})$ and
$\bm{q}_\alpha^{\rm T}=(q_{1,\alpha},q_{2,\alpha},\cdots)$.
The remaining actions for the bath and the interaction are given respectively by
 \begin{align}
S_{\rm B}[\{\bm{q}_\alpha\}]=\sum_{\alpha\in\mathcal{B}}\sum_{n}& \int_0^{\beta\hbar} d\tau\; \Big[ \frac {m_{n,\alpha}} 2 
\left(\frac{dq_{n,\alpha}}{d\tau}\right)^2 \nonumber \\
&+\frac 1 2 m_{n,\alpha} \omega^2_{n,\alpha} q^2_{n,\alpha}(\tau)\Big], 
\end{align}
and
\begin{align}
 &S_{\rm I}[\bm{x},\{\bm{q}_\alpha\}] \\
 &=\sum_{\alpha\in\mathcal{B}} \int_0^{\beta\hbar} d\tau\; \left[ -\sum_n \kappa_{n,\alpha} q_{n,\alpha}(\tau)x_\alpha(\tau)
+ \frac{\mu_\alpha }{2} x_\alpha^2(\tau)\right] \nonumber 
 \end{align}
 with
 \begin{align}
\mu_\alpha\equiv\sum_n\frac{\kappa^2_{n,\alpha}}{m_{n,\alpha}\omega^2_{n,\alpha}}=2\int_0^\infty d\omega\; \frac{J_\alpha(\omega)}{\omega}.
\label{mua}
\end{align}

The matrix element of the MFG state is obtained by tracing over the bath variables. 
This type of path integral is well-known in the study of the quantum Brownian motion \cite{grabert_quantum_1988}. 
The path integral over the bath variables
results in the so-called Feynman-Vernon influence functional \cite{Feynman_1963, grabert_quantum_1988,QDS_Weiss}.
In our case, we can easily modify it to the imaginary time path integral.
As a result, we have the influence functional
obtained by tracing out the $\alpha$-th bath as
\begin{align}
\Psi_\alpha[x_\alpha] =
& -\int_0^{\beta\hbar} d\tau \int_0^\tau d\tau'\; K_\alpha(\tau-\tau')x_\alpha(\tau)x_\alpha(\tau') \nonumber  \\
&   +\frac{\mu_\alpha}{2}\int_0^{\beta\hbar}d\tau\; x_\alpha^2(\tau),
\label{influence0}
\end{align}
where
\begin{align}
K_\alpha(\tau)\equiv\int_0^\infty d\omega\; J_\alpha(\omega)
 \frac{\cosh(\frac 1 2\beta\hbar\omega-\omega \tau)}{\sinh(\frac 1 2\beta\hbar\omega)}.
 \label{Ka}
 \end{align}
Now the matrix element of the MFG state with respect to the system position variables is given by 
\begin{align}
&\langle \bm{\bar{x}}\vert \hat{\rho}_{\rm mfG} \vert \bm{\bar{x}}'\rangle 
=\int \prod_{\alpha\in\mathcal{B}}\prod_n d\bar{q}_{n,\alpha}\; \langle \bm{\bar{x}}, \{\bm{\bar{q}}_\alpha\} \vert
 \hat{\rho}_\beta \vert \bm{\bar{x}}', \{\bm{\bar{q}}_\alpha\} \rangle   \nonumber \\
 =& \frac 1 Z \int_{\bm{x}(0)=\bm{\bar{x}'}}^{\bm{x}(\beta\hbar)=\bm{\bar{x}}} \mathcal{D}\bm{x}(\tau)
 \exp\left [-\frac 1 {\hbar}( S_{\rm S}[\bm{x}]+\sum_{\alpha\in\mathcal{B}}\Psi_\alpha[x_\alpha])\right],  \label{mfg_path_mul}
\end{align}
where $Z=Z_{\rm B}/Z_\beta$ with $Z_{\rm B}=\mathrm{Tr}_{\rm B} \exp(-\beta H_{\rm B})$.
We note that we can rewrite Eq.~(\ref{influence0}) as (see Appendix \ref{app:1})
 \begin{align}
\Psi_\alpha[x_\alpha] = \frac M 2\int_0^{\beta\hbar} d\tau \int_0^{\beta\hbar} d\tau'\; L_\alpha(\tau-\tau')x_\alpha(\tau)x_\alpha(\tau'),
\label{influence}
\end{align}
where $L_\alpha(\tau)$ is defined as a Fourier series expansion in the interval $0\le \tau\le \beta\hbar$ as
\begin{align}
L_\alpha(\tau)=\frac{1}{\beta\hbar}\sum_{r=-\infty}^{\infty}\zeta^{(\alpha)}_r e^{i\nu_r\tau}.
\label{La}
\end{align}
with the Matsubara frequency $\nu_r=2\pi r/(\beta\hbar)$ and
\begin{align}
\zeta^{(\alpha)}_r=\frac 1 M \int_0^\infty d\omega\; \frac{J_\alpha(\omega)}{\omega} \frac{2\nu^2_r}{\omega^2+\nu^2_r}.
\label{zetara}
\end{align}

Since the path integral in Eq.~(\ref{mfg_path_mul}) is Gaussian, it can be evaluated by considering the stationary path
and the fluctuation around it. The path integral over the fluctuation will give a constant independent of the endpoints,
which can be determined later
from the normalization of $\hat{\rho}_{\rm mfG}$ \cite{grabert_quantum_1988}. The stationary path is determined from
\begin{align}
0=&-M \frac{d^2 x_i(\tau)}{d\tau^2}+\sum_{j=0}^{N-1}\Lambda_{ij}x_j(\tau) \nonumber \\
&+M\sum_{\alpha\in\mathcal{B}}\int_0^{\beta\hbar}d\tau'\; L_\alpha(\tau-\tau')x_\alpha(\tau').
\label{stationary_mul}
\end{align}
for $i=0,1,\cdots,N-1$. This is to be solved with the boundary conditions
$\bm{x}(0)=\bm{\bar{x}'}$ and $\bm{x}(\beta\hbar)=\bm{\bar{x}}$.
The solution can be conveniently obtained in terms of the Fourier series expansion
\begin{align}
x_i(\tau)=\frac 1 {\beta\hbar}\sum_{r=-\infty}^\infty x_{i,r} e^{i\nu_r \tau}.
\label{xexpand_mul}
\end{align}
The detailed procedure is given in Appendix \ref{app:2}. The solution
for the Fourier modes $x_{i,r}$ in a matrix form are given by
\begin{align}
\bm{x}_r=i\nu_r  \bm{G}(\nu_r) (\bm{\bar{x}}-\bm{\bar{x}'}) +\frac 1 2 \bm{G}(\nu_r) \bm{F}^{-1}\left(\bm{\bar{x}}+\bm{\bar{x}'}\right),
\label{xr_mul}
\end{align}
where the matrices 
\begin{align}
\bm{G}(\nu_r)\equiv \left[  \nu_r^2 \bm{1} +\frac 1 M \bm{\Lambda}+ \bm{\Sigma}(\nu_r) \right]^{-1},
\label{green_mul}
\end{align}
and
\begin{align}
\bm{F}\equiv \frac 1 {\beta\hbar}\sum_{r=-\infty}^\infty \bm{G}(\nu_r).
\label{F_mul}
\end{align}
Here $\bm{\Sigma}(\nu_r)$
is a diagonal matrix  whose diagonal element $\Sigma_{jj}$ is equal to 
$\zeta_r^{(\alpha)}$ given in Eq.~(\ref{zetara}) if $j$ belongs to the set $\mathcal{B}$ and is equal to zero otherwise.

If we insert the solution Eq.~(\ref{xr_mul}) with Eq.~(\ref{xexpand_mul}) into the expression of the MFG state, 
Eq.~(\ref{mfg_path_mul}),
we obtain (see Appendix \ref{app:3})
\begin{align}
&\langle \bm{\bar{x}}\vert \hat{\rho}_{\rm mfG} \vert \bm{\bar{x}'}\rangle 
=C\exp\Bigg[ -\frac{M}{2\hbar} (\bm{\bar{x}}-\bm{\bar{x}'})^{\rm T} \bm{A} \; (\bm{\bar{x}}-\bm{\bar{x}'}) \nonumber \\
&~~~~~~~~~~~~~~~ -\frac{M}{2\hbar} \left(\frac{ \bm{\bar{x}}+\bm{\bar{x}'} }{2}\right)^{\rm T} \bm{F}^{-1} \;
\left(\frac{ \bm{\bar{x}}+\bm{\bar{x}'} }{2}\right) \Bigg],
\label{mfg_element_mul}
\end{align}
where
\begin{align}
\bm{A}\equiv \frac {1}{\beta\hbar}\sum_{r=-\infty}^\infty \left[  \bm{1} - \nu^2_r \bm{G}(\nu_r)\right] .
\end{align}
The overall constant $C$ is determined from the normalization
\begin{align}
1=\int d^N \bm{\bar{x}}\; \langle \bm{\bar{x}}\vert \rho_{\rm mfG} \vert \bm{\bar{x}}\rangle,
\end{align}
which gives
\begin{align}
C=\left( \frac{M}{2\pi\hbar} \right)^{\frac N 2} 
\frac{1}{(\det\bm{F})^{\frac 1 2}}
\end{align}

The MFG state is Gaussian and characterized by the following covariance matrices,
which can easily be calculated as
\begin{align}
\langle \hat{x}_i \hat{x}_j\rangle_{\rm mfG} = \mathrm{Tr}[ \hat{x}_i \hat{x}_j \hat{\rho}_{\rm mfG}] = \frac{\hbar}{M} F_{ij},
\label{xx_mul}
\end{align}
\begin{align}
\langle \hat{p}_i \hat{p}_j \rangle_{\rm mfG} = \mathrm{Tr}[ \hat{p}_i \hat{p}_j \hat{\rho}_{\rm mfG}] = M \hbar A_{ij},
\label{pp_mul}
\end{align}
and
\begin{align}
\langle\{ \hat{x}_i ,\hat{p}_j\}\rangle_{\rm mfG} =\mathrm{Tr}[(\hat{x}_i \hat{p}_j +\hat{p}_j \hat{x}_i )\hat{\rho}_{\rm mfG}]=0.
\end{align}

\section{System in Contact with a single heat bath}
\label{sec:single}

In this section, using the general formalism developed above, we develop a nonperturbative method
to evaluate matrix Green's function in Eq.~(\ref{green_mul}). 
To do this,
we first focus on coupled harmonic oscillators one end of which is in contact with 
a heat bath at inverse temperature $\beta$. In the notation of Eq.~(\ref{HBI}), we take $\mathcal{B}=\{0\}$, that is
the $\alpha=0$ oscillator is interacting with the heat bath. Then the only non-vanishing 
matrix element of $\bm{\Sigma}(\nu_r)$ which appears in Eq.~(\ref{green_mul})
is given by $\Sigma_{00}(\nu_r)=\zeta_r$.  Here we have dropped the index $\alpha$ in the definition
of $\zeta_r$ in Eq.~(\ref{zetara}) as there is only one bath.

We note that the calculation of the MFG state comes down to evaluating the Green's function 
$\bm{G}$ in Eq.~(\ref{green_mul}).
A convenient way to do this to express $\bm{G}$ in terms of 
$\bm{G}^{(0)}(\nu_r)\equiv [\nu^2_r \bm{1}+\bm{\Lambda}/M]^{-1}$,
which is the Green's function in the absence of coupling to the bath.
This is a standard technique used to calculate the full Green's function perturbatively. See for example Ref.~\cite{economou2006green}.
If we use $\bm{G}^{(0)}$ in Eq.~(\ref{mfg_element_mul}), then we can get the
matrix element of the system Gibbs state $\hat{\rho}_{\rm G}$.
We note that $\bm{G}^{(0)}$ can be obtained by diagonalizing $\bm{\Lambda}$.
Once we have $\bm{G}^{(0)}$, we can write
\begin{align}
\bm{G}(\nu_r)=& \left[ (\bm{G}^{(0)})^{-1}(\nu_r)+ \bm{\Sigma}(\nu_r)\right]^{-1}  \nonumber \\
=& \left[ \bm{1}+\bm{G}^{(0)} \bm{\Sigma}\right]^{-1} \bm{G}^{(0)} \label{green_exp} \\
=&\left[\bm{1}- \bm{G}^{(0)} \bm{\Sigma}
+ \bm{G}^{(0)} \bm{\Sigma}\bm{G}^{(0)} \bm{\Sigma}+\cdots \right] \bm{G}^{(0)} . \nonumber
\end{align}
Since the only non-vanishing element of $\bm{\Sigma}$ is $\Sigma_{00}=\zeta_r$, we can write 
\begin{align}
G_{ij}(\nu_r)=&G^{(0)}_{ij} -\zeta_r G^{(0)}_{i0}G^{(0)}_{0j}
+\zeta^2_r G^{(0)}_{i0}G^{(0)}_{00}G^{(0)}_{0j} \nonumber \\
&-\zeta^3_r G^{(0)}_{i0}[G^{(0)}_{00}]^2 G^{(0)}_{0j} +\cdots .
\end{align}
By summing the infinite series, we thus have
\begin{align}
\bm{G}(\nu_r)= \bm{G}^{(0)}(\nu_r) +\Delta \bm{G}(\nu_r),
\label{green_result}
\end{align}
where
\begin{align}
\Delta G_{ij}(\nu_r)\equiv -\zeta_r \frac{G^{(0)}_{i0}(\nu_r)G^{(0)}_{0j}(\nu_r) }{ 1+\zeta_r G^{(0)}_{00} (\nu_r)}.
\label{delta_g}
\end{align}
For later purpose, we rewrite Eq.~(\ref{delta_g}) as
\begin{align}
\Delta\bm{G}(\nu_r)=-  \bm{G}^{(0)}(\nu_r) \bm{\Gamma}(\nu_r) \bm{G}^{(0)}(\nu_r)
\label{deltag1}
\end{align}
using the matrix $\bm{\Gamma}$ with only one nonvanishing element defined by
\begin{align}
\Gamma_{ij}(\nu_r)= \delta_{i0}\delta_{j0}\frac{\zeta_r}{1+\zeta_r G^{(0)}_{00} (\nu_r)}.
\label{gamma1}
\end{align}

Another way of writing the full Green's function is 
\begin{align}
&G_{ij}(\nu_r)  \label{full_green} \\
&=\frac{
G^{(0)}_{ij}(\nu_r)+\zeta_r\{ G^{(0)}_{ij}(\nu_r)G^{(0)}_{00}(\nu_r)-G^{(0)}_{i0}(\nu_r)G^{(0)}_{0j}(\nu_r) \} }
{1+\zeta_r G^{(0)}_{00}(\nu_r)}. \nonumber
\end{align}
We can then use this equation to obtain the expressions for 
the covariances with respect to the MFG state in Eqs~(\ref{xx_mul}) and (\ref{pp_mul}) as
\begin{widetext}
\begin{align}
\langle \hat{x}_i \hat{x}_j\rangle_{\rm mfG}=\frac{1}{M\beta}\left[ G^{(0)}_{ij}(0)
+2\sum_{r=1}^{\infty} \frac{
G^{(0)}_{ij}(\nu_r)+\zeta_r\{ G^{(0)}_{ij}(\nu_r)G^{(0)}_{00}(\nu_r)-G^{(0)}_{i0}(\nu_r)G^{(0)}_{0j}(\nu_r) \} }
{1+\zeta_r G^{(0)}_{00}(\nu_r)} \right]
\label{xx_full}
\end{align}
and
\begin{align}
\langle \hat{p}_i \hat{p}_j\rangle_{\rm mfG}=\frac{M}{\beta}\left[  \delta_{ij}
+2 \sum_{r=1}^{\infty} \left\{  \delta_{ij}-\frac{
\nu^2_r G^{(0)}_{ij}(\nu_r)+\nu^2_r\zeta_r\{ G^{(0)}_{ij}(\nu_r)G^{(0)}_{00}(\nu_r)-G^{(0)}_{i0}(\nu_r)G^{(0)}_{0j}(\nu_r) \} }
{1+\zeta_r G^{(0)}_{00}(\nu_r)} \right\} \right],
\label{pp_full}
\end{align}
where the summation is now over positive frequencies $\nu_r$.
\end{widetext}

\subsection{Example: Chain of Harmonic Oscillators in Contact with a Heat Bath}
\label{sec:chain1}

\begin{figure}
\resizebox{0.95\columnwidth}{!}{\includegraphics{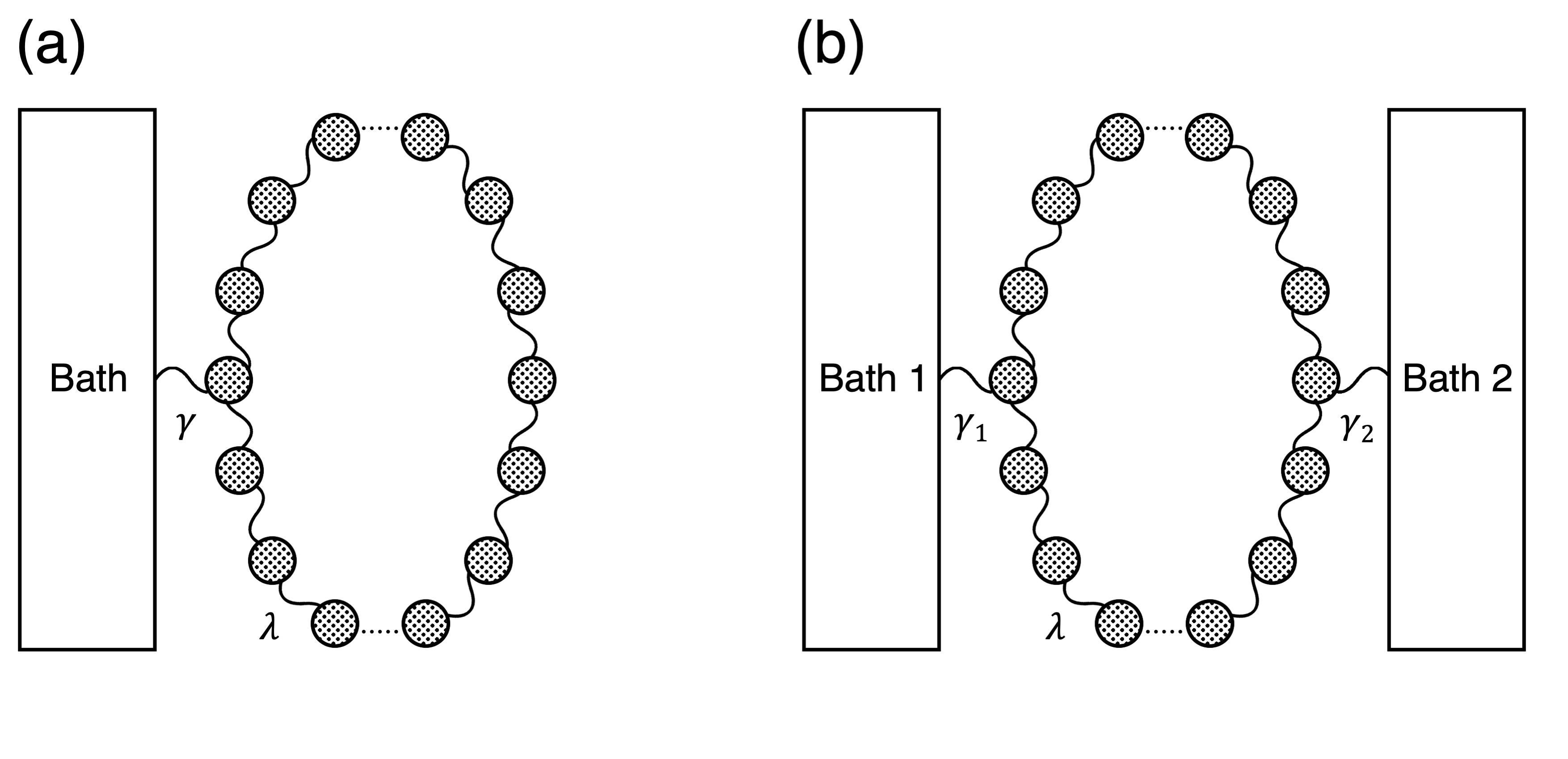}}
\caption{Schematic diagrams depicting a chain of harmonic oscillators interacting with (a) a single heat bath
and (b) two heat baths.}
\label{fig:osc}
\end{figure}

In this subsection, we apply the method developed in the previous section to a chain of coupled 
oscillators one end of which is in contact with a heat bath. We consider 
a system described by Hamiltonian
\begin{align}
\hat{H}_{\rm S}=\sum_{i=0}^{N-1}\left( 
\frac{\hat{p}_i^2}{2M}+\frac 1 2 M\Omega^2 \hat{x}^2_i
-M\lambda \hat{x}_i\hat{x}_{i+1} \right).
\end{align}
The parameter $\lambda$ describes the coupling between the neighboring oscillators.
For convenience, we assume that the system satisfies the periodic boundary condition,
$\hat{x}_N=\hat{x}_0$. The oscillators thus
form a ring (see Fig.~\ref{fig:osc} (a) for schematic description). 
The matrix $\bm{\Lambda}$ in Eq.~(\ref{Hs}) is given by
\begin{align}
\bm{\Lambda}=M\Omega^2\bm{1}-M\lambda \bm{V},
\label{lambdav}
\end{align}
where
\begin{align}
\bm{V}=
\begin{pmatrix}
0 & 1 & 0 &0 & \cdots & 0 & 1 \\
1 & 0 & 1  &0 & \cdots & 0  & 0\\
0 & 1 & 0  &1 & \cdots & 0  & 0\\
\vdots & \vdots & \vdots& \vdots &  & \vdots & \vdots\\
1 & 0 & 0& 0 &\cdots &1 & 0
\end{pmatrix}
\label{vmatrix}
\end{align}
As shown in Appendix \ref{app:4},
the eigenvalues of $\bm{\Lambda}$ are given by $M\Omega^2_k$, $k=0,1,\cdots,N-1$, where
\begin{align}
\Omega^2_k \equiv \Omega^2 - 2\lambda\cos\left(\frac{2\pi}{N}k\right).
\label{evalue}
\end{align}
Note that there are two-fold degeneracies as $\Omega_{N-k}=\Omega_{k}$.
For stability we require $2\vert \lambda \vert <\Omega^2 $. The eigenvectors of $\bm{\Lambda}$ can be used 
to construct an orthogonal matrix $\bm{R}$, which diagonalizes $\bm{\Lambda}$ as
$\bm{R}^{\rm T}\bm{\Lambda}\bm{R}=\mathrm{diag}(M\Omega^2_0,\cdots,M\Omega^2_{N-1})$.
We therefore have
\begin{align}
&\bm{G}^{(0)}(z)=\left[  z^2\bm{1}+\frac 1 M \bm{\Lambda} \right]^{-1} \label{matrixR} \\
&~~~~~ =\bm{R} ~\mathrm{diag}\left(\frac{1}{z^2+\Omega^2_0},\cdots,\frac{1}{z^2+\Omega^2_{N-1}}\right)
\bm{R}^{\rm T}. \nonumber 
\end{align}
In Appendix \ref{app:4}, we present explicit eigenvectors and consequently $\bm{R}$ which diagonalizes $\bm{V}$ and $\bm{\Lambda}$.

Now the full Green's function $\bm{G}$ in the presence of the coupling to the bath 
can be obtained using Eq.~(\ref{green_result}). 
The effect of coupling to the bath is characterized by $\zeta^{(\alpha)}_r$ in Eq.~(\ref{zetara}) and the bath spectral density 
$J_\alpha(\omega)$ in Eq.~(\ref{Ja}).
We will again drop the index $\alpha$ as there is only one bath.
We consider an Ohmic bath with the cutoff $\omega_D$ and the coupling strength $\gamma$, which is given by
\begin{align}
J(\omega)=\frac{2M\gamma}{\pi}\omega\frac{\omega^2_D}{\omega^2+\omega^2_D}. \label{spectral}
\end{align}
Putting this into Eq.~(\ref{zetara}), we have
\begin{align}
\zeta_r =2\gamma \frac{\omega_D \vert \nu_r \vert }{\vert \nu_r \vert+\omega_D}. \label{ohmic}
\end{align}

In the following, we present explicit expressions
for the MFG state for general values of $N\ge 2$. As we have seen above  in Eqs.~(\ref{xx_mul}) and (\ref{pp_mul}), the MFG state
is completely characterized by its covariances. 
First, using the eigenvectors obtained in Appendix \ref{app:4} and Eq.~(\ref{matrixR}), we can write 
\begin{align}
G^{(0)}_{ij}(z)=\frac 1 N \sum_{k=0}^{N-1}\frac{\cos\left(\frac{2\pi}{N}(i-j)k\right)}{z^2+\Omega^2_k} .
\label{G0}
\end{align}
We then obtain the covariances with respect to the
system Gibbs state $\hat{\rho}_{\rm G}$, which we denote by $\langle \cdots \rangle_{\rm G}$. 
This can be done by 
using Eqs.~(\ref{xx_mul}) and (\ref{pp_mul}) with $\bm{G}$ replaced by $\bm{G}^{(0)}$. Using the summation formula
\begin{align}
\sum_{r=-\infty}^\infty \frac 1{\nu^2_r+\omega^2}=\left(\frac{\beta\hbar}{2\omega}\right)\coth\left(\frac{\beta\hbar\omega}{2}\right),
\label{summation}
\end{align}
we obtain 
\begin{align}
\langle \hat{x}_i \hat{x}_j\rangle_{\rm G} = \frac {\hbar} {2MN} 
    \sum_{k=0}^{N-1} \frac{\coth\left(\frac{\beta\hbar\Omega_k}{2}\right)}{\Omega_k} 
\cos\left(\frac{2\pi}{N}(i-j)k\right) .
\label{xij_g}
\end{align}
For the momentum covariances, we first note from Eq.~(\ref{matrixR}) that 
\begin{align}
&\bm{1}-z^2\bm{G}^{(0)}(z) \nonumber  \\
&=\bm{R} ~\mathrm{diag}\left(\frac{\Omega^2_0}{z^2+\Omega^2_0},\cdots,\frac{\Omega^2_{N-1}}{z^2+\Omega^2_{N-1}}\right)
\bm{R}^{\rm T}.
\end{align}
We have from Eq.~(\ref{pp_mul})
\begin{align}
\langle \hat{p}_i \hat{p}_j\rangle_{\rm G} = \frac {M\hbar}{2N}
  \sum_{k=0}^{N-1} \Omega_k \coth\left(\frac{\beta\hbar\Omega_k}{2}\right)
\cos\left(\frac{2\pi}{N}(i-j)k\right).
\label{pij_g}
\end{align}

Now the covariances  with respect to the MFG state can be obtained from the full Green's function
in Eq.~(\ref{green_result}) with Eq.~(\ref{delta_g}). We therefore have
\begin{align}
&\langle \hat{x}_i \hat{x}_j\rangle_{\rm mfG} =\langle \hat{x}_i \hat{x}_j\rangle_{\rm G} +\Delta\langle \hat{x}_i \hat{x}_j\rangle, \label{deltaxx_def}\\
&\langle \hat{p}_i \hat{p}_j\rangle_{\rm mfG} =\langle \hat{p}_i \hat{p}_j\rangle_{\rm G} +\Delta\langle \hat{p}_i \hat{p}_j\rangle \label{deltapp_def},
\end{align}
where the differences compared to those obtained with respect to the system Gibbs state are given in terms of $\Delta\bm{G}$ in
Eq.~(\ref{delta_g}) by
\begin{align}
&\Delta\langle \hat{x}_i \hat{x}_j\rangle =\frac 1 {M\beta} \sum_{r=-\infty}^\infty \Delta G_{ij}(\nu_r) \label{deltaxx}\\
&\Delta\langle \hat{p}_i \hat{p}_j\rangle =-\frac{M}{\beta}  \sum_{r=-\infty}^\infty \nu^2_r \Delta G_{ij}(\nu_r). \label{deltapp}
\end{align}

\begin{figure}
	\resizebox{0.95\columnwidth}{!}{\includegraphics{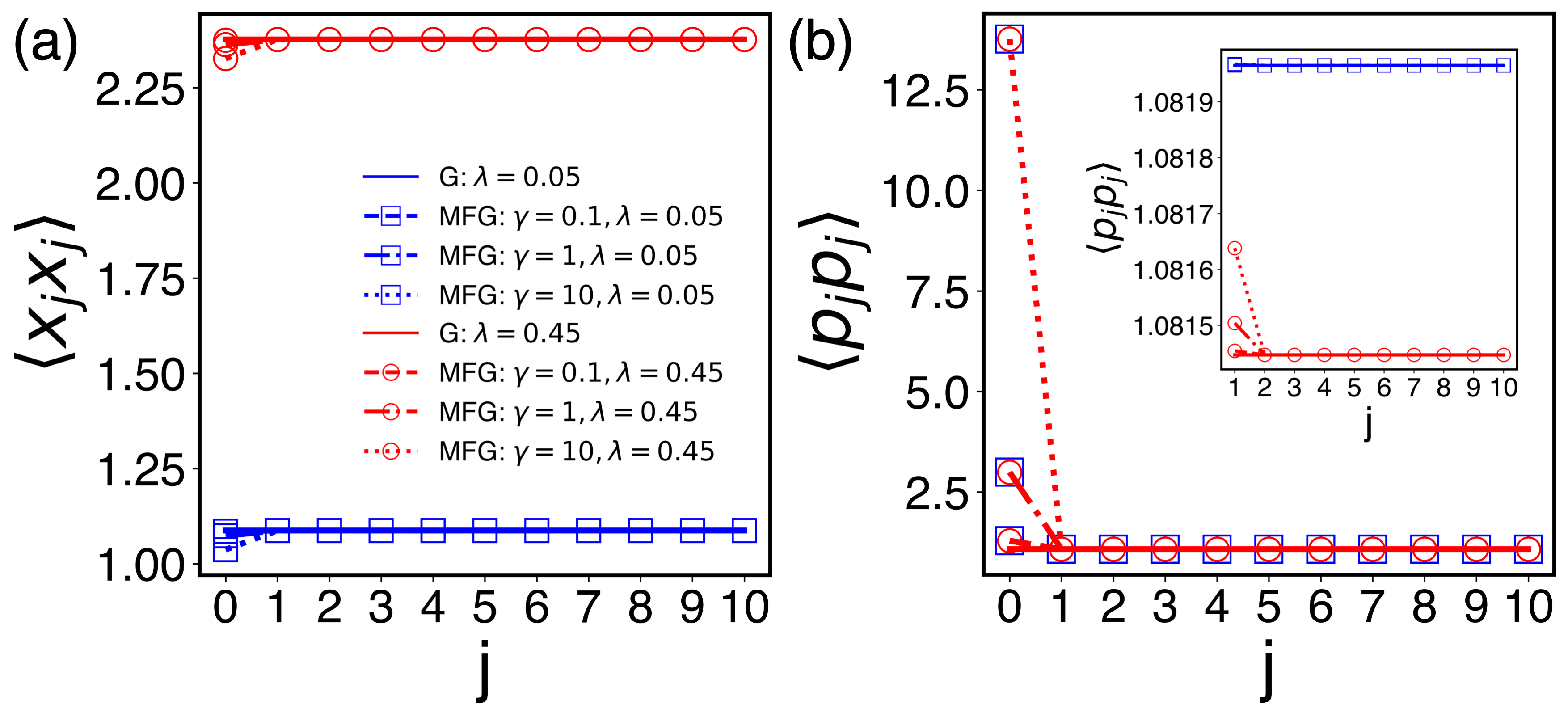}}
	\caption{The covariances (a) $\langle\hat{x}^2_j\rangle$ and (b) $\langle\hat{p}^2_j\rangle$ as functions of the position $j$ of the  $N=20$ oscillators. 
	At $j=0$, the system is in contact with the heat bath maintained at the inverse temperature $\beta=1$. 
	The solid horizontal lines are
	those calculated with respect to the system Gibbs (G) states. The data points and the broken lines are obtained from the averages 
	with respect to the MFG states for various values of  the inter-oscillator coupling $\lambda=0.05$ (squares) and $\lambda=0.45$ (circles)
	and the coupling strength to the bath $\gamma=0.1,1$ and 10.
	The inset in (b) is the same plot excluding $j=0$ point. }
	\label{fig:b1beta1}
\end{figure}

\begin{figure}
	\resizebox{0.95\columnwidth}{!}{\includegraphics{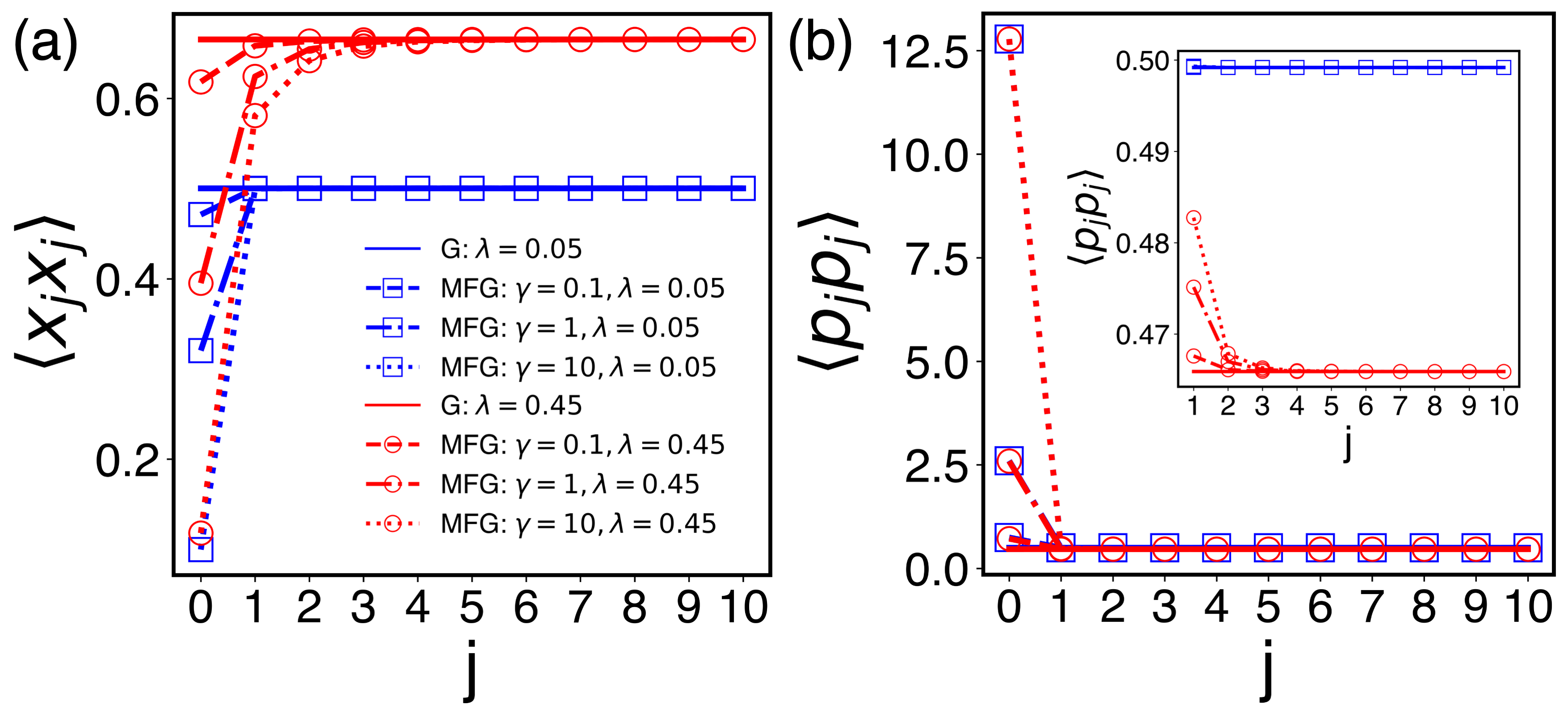}}
	\caption{ The covariances (a) $\langle\hat{x}^2_j\rangle$ and (b) $\langle\hat{p}^2_j\rangle$ as functions of the position $j$ of the oscillators at
	the inverse temperature $\beta=100$. The other parameters are the same as in Fig.~\ref{fig:b1beta1}. }
	\label{fig:b1beta100}
\end{figure}

We evaluate the infinite sums in Eqs.~(\ref{deltaxx}) and (\ref{deltapp}) numerically and calculate the covariances 
with respect to the MFG state. For the numerical calculations, we present $\langle \hat{x}_i \hat{x}_j\rangle$
and $\langle \hat{p}_i \hat{p}_j\rangle$ in units of $\hbar/(M\Omega)$ and $\hbar M\Omega$, respectively.
In Figs.~\ref{fig:b1beta1} and \ref{fig:b1beta100}, we show the results for $\langle \hat{x}^2_i \rangle$
and $\langle \hat{p}^2_i \rangle$ calculated with respect to the MFG state and compare them
with those obtained from the system Gibbs state for various values of the coupling strength $\gamma$
(in the unit of $\Omega$) to the bath and of the inter-oscillator coupling constant $\lambda$ (in the unit of $\Omega^2$).
The calculations were done for a ring of $N=20$ coupled oscillators where $j=0$ oscillator is coupled with the heat bath
at inverse temperature $\beta$. We take the Ohmic bath described by Eqs.~(\ref{spectral}) and (\ref{ohmic}) with
$\omega_D=100$ (in the unit of $\Omega$).

At relatively high temperature, $\beta=1$  (The inverse temperature is in the unit of $1/(\hbar\Omega)$.), 
Fig.~\ref{fig:b1beta1} (a) shows that $\langle\hat{x}^2_j \rangle_{\rm mfG}$ exhibits very little variation compared to those 
obtained from the system Gibbs state, which is independent of $j$ (see Eq.~(\ref{xij_g})). The small difference between the 
MFG and the system Gibbs averages occurs only near the contact point $j=0$ between the system and the bath.
On the other hand,
the momentum covariance $\langle \hat{p}^2_j\rangle_{\rm mfG}$ shows a different behavior as one can see
from Fig.~\ref{fig:b1beta1} (b). The value $\langle \hat{p}^2_0\rangle_{\rm mfG}$ right at the point of contact with the bath 
increases quite rapidly as the coupling strength $\gamma$ to the bath increases. As we will see below in the next
subsection, this is a characteristic behavior of a quantum system where the coupling between the {\it position} operators of
the system and the heat bath gets stronger. As we can see in the inset of the same figure, away from the
point of contact, the effect of the coupling to the bath quickly disappears as in the position covariance.

\begin{figure}
	\resizebox{0.95\columnwidth}{!}{\includegraphics{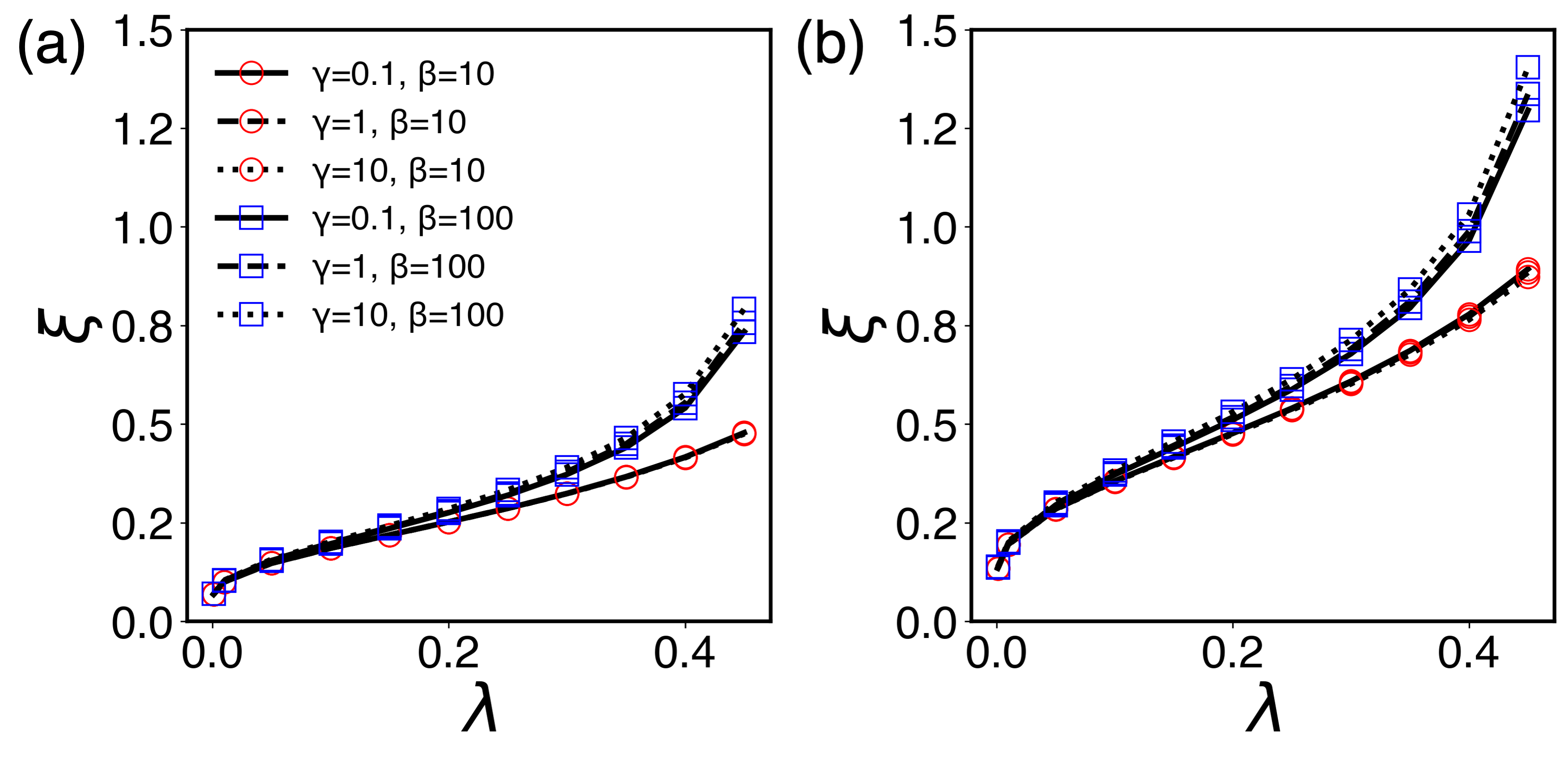}}
	\caption{The skin depth $\xi$ obtained from the exponential curve fitting of (a) $\vert \Delta \langle \hat{x}^2_j\rangle\vert$ 
	and (b) $\vert \Delta \langle \hat{x}_0\hat{x}_j\rangle\vert$  (see main text) as a function of the 
	inter-oscillator coupling  $\lambda$ for the coupling strength to the bath $\gamma=0.1$, 1 and 10, and for
	the inverse temperature $\beta=10$ (circles) and 100 (squares).}
	\label{fig:corr}
\end{figure}

At lower temperatures, the effect of the coupling to the bath in $\langle\hat{x}^2_j\rangle_{\rm mfG}$ is more prominent 
as we can see in Fig.~\ref{fig:b1beta100} (a) at the inverse temperature $\beta=100$.
The effect, however, disappears as we move away from the point of contact with the bath. In fact, at lower temperatures $\beta\gtrsim 10$,
we find that the difference defined in Eq.~(\ref{deltaxx_def}) is well described 
by an exponential decay $\vert \Delta \langle \hat{x}^2_j\rangle\vert\sim\exp(-j/\xi)$
with the characteristic length $\xi$. This is quite similar to the skin effect found in Ref.~\cite{burke_structure_2024},
where the effect of coupling to the environment decays exponentially with the distance from the system-environment boundary.
In Fig.~\ref{fig:corr} (a), we plot the skin depth $\xi$ as a function of the
inter-oscillator coupling $\lambda$ for various values of the temperature and the coupling strength $\gamma$ to the bath. 
We note that $\xi$ is quite small ($\lesssim 1$)
even at low temperatures and strong system-bath couplings, and that it shows only a moderate increase as a function of $\lambda$
and $\beta$. A somewhat surprising result is that the effect of the bath measured in terms of $\xi$ is almost independent of the 
coupling strength $\gamma$ to the bath for given values of $\beta$ and $\lambda$.
The momentum covariance $\langle\hat{p}^2_j\rangle_{\rm mfG}$ at lower temperatures, on the other hand, behaves quite similarly to 
that at higher temperatures as can be seen from Fig.~\ref{fig:b1beta100} (b). The main feature still is a sharp increase 
of $\langle \hat{p}^2_0\rangle_{\rm mfG}$ as $\gamma$ increases. The effect of the coupling to the bath quickly disappears 
as can be seen in the inset of Fig.~\ref{fig:b1beta100} (b).

\begin{figure}
	\resizebox{0.95\columnwidth}{!}{\includegraphics{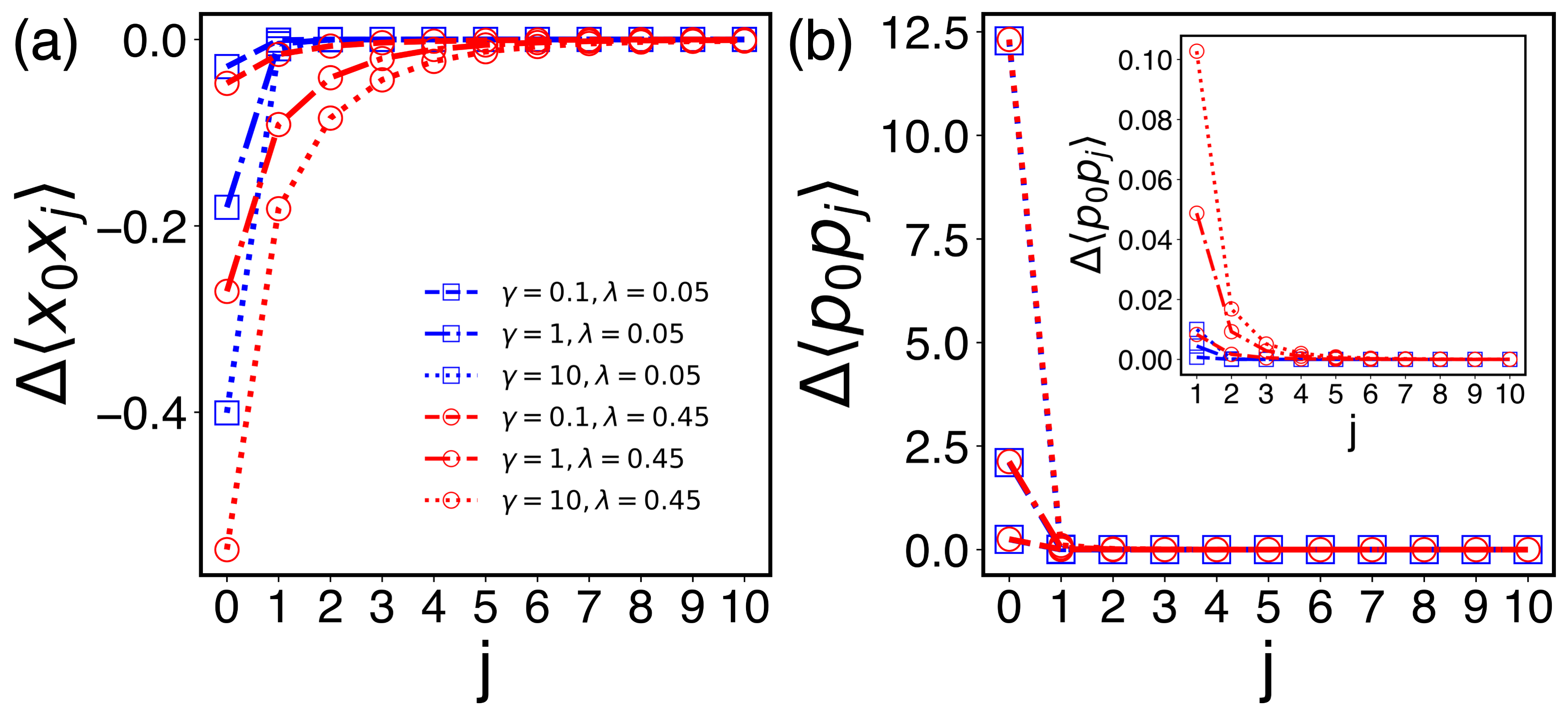}}
	\caption{ The differences (a) $\Delta\langle\hat{x}_0 \hat{x}_j\rangle$ and (b) $\Delta\langle\hat{p}_0\hat{p}_j\rangle$  of covariances
	between those calculated with respect to the Gibbs and the MFG states as functions of the position $j$ of the $N=20$ oscillators for various values of the coupling strength $\gamma=0.1, 1$ and 10 to the bath and the inter-oscillator coupling $\lambda=0.05$ (squares) and 0.45 (circles). The system is in contact with the heat bath at $j=0$.
	The inverse temperature is $\beta=100$. The inset in (b) shows the same data from $j=1$. }
	\label{fig:delta}
\end{figure}

In order to study the effect of the coupling to the bath on the MFG state further, we look at another type of covariances,
$\langle \hat{x}_0\hat{x}_j\rangle_{\rm mfG}$ and $\langle \hat{p}_0\hat{p}_j\rangle_{\rm mfG}$, which connect operators at
different sites, and compare them with those obtained with respect to the system Gibbs state. 
We again study the effect of the coupling to the bath using the differences, $\Delta\langle\hat{x}_0\hat{x}_j\rangle$ 
and $\Delta\langle \hat{p}_0\hat{p}_j\rangle$ between the MFG and the system Gibbs states, which are plotted in 
Fig.~\ref{fig:delta} at inverse temperature $\beta=100$ for $N=20$ oscillators. 
We note that in this case the system Gibbs averages as well as the MFG ones show a spatial dependence on $j$ as one can see from 
Eqs.~(\ref{xij_g}) and (\ref{pij_g}). The differences between them, however, show a similar behavior to the previous case
of $\Delta\langle \hat{x}_j^2\rangle$ as can be seen from Fig.~\ref{fig:delta} (a). In fact, we find that
$\vert\Delta\langle\hat{x}_0\hat{x}_j\rangle\vert$ is also well described by an exponential decay $\sim\exp(-j/\xi)$. 
We plot the skin depth $\xi$ for this covariance in Fig.~\ref{fig:corr} (b) for various values of the parameters.
We note that the skin depths in this case are slightly bigger than those for $\Delta\langle \hat{x}_j^2\rangle$, but
the overall behavior of $\xi$ as a function of the coupling constants and temperatures 
is quite similar to the previous case.
The main point we make here is that the effect of the heat bath on the equilibrium state  measured with $\xi$
in both cases is quite short-ranged reaching only over one site.
The difference in the momentum covariance $\Delta\langle \hat{p}_0\hat{p}_j\rangle$ connecting different sites also 
shows a similar behavior to $\Delta\langle \hat{p}^2_j\rangle$. Apart from the increasing 
$\Delta\langle \hat{p}^2_0\rangle_{\rm mfG}$ at $j=0$ as $\gamma$ gets bigger, the effect of the coupling to the bath 
quickly disappears as we move away from the contact point as we can see in Fig.~\ref{fig:delta} (b).

\subsection{The Ultra Strong Coupling (USC) Limit}
In  this subsection, we investigate the MFG state obtained above, which is an exact result,
in the limit when the coupling between
the system and the bath is arbitrarily large.  This is to make a connection with the ultra strong coupling (USC) limit 
obtained recently in Ref.~\cite{cresser_weak_2021} of the MFG state for
a general system coupled with a bath. 
If the interaction Hamiltonian between system and bath is given by $\kappa \hat{X} \hat{B}$
with a system operator $\hat{X}$ and a bath operator $\hat{B}$, and  the coupling strength $\kappa$,
it was shown in Ref.~\cite{cresser_weak_2021} that the MFG state in the limit $\kappa\to\infty$
is diagonal in the eigenstates of $\hat{X}$. More explicitly, the ultra strong coupling limit
is given by \cite{cresser_weak_2021}
\begin{align}
\hat{\rho}_{\rm USC}\sim \exp[-\beta \sum_n \hat{P}_n \hat{H}_{\rm S} \hat{P}_n ],
\label{usc}
\end{align}
where $\hat{P}_n$ is the projection operator on the eigenstates $\vert X_n \rangle$ of $\hat{X}$ with eigenvalue $X_n$. 
In other words, the USC limit is characterized by the effective Hamiltonian of mean force
$H_{\rm USC}\equiv \sum_n \hat{P}_n \hat{H}_{\rm S} \hat{P}_{n}$, which is just the system Hamiltonian but 
projected onto the eigenstates of the system operator taking part in the coupling to the bath.

The reason we study our present system in this limit is twofold. First, 
for the case of harmonic oscillators, the system operator coupled to the bath is just the position operator $\hat{x}_0$ 
at $j=0$ thus has a continuous spectrum.
Therefore, the USC limit of the exact MFG state for the quantum harmonic oscillators can provide
a generalization of Eq.~(\ref{usc})  which was originally obtained in Ref.~\cite{cresser_weak_2021} for 
the coupling system operator with a discrete spectrum
to a continuous one. 

To illustrate this point, we consider the simplest case of a single ($N=1$) damped harmonic oscillator 
\cite{Grabert_Weiss_Talkner_1984, hilt_hamiltonian_2011} in the USC limit.  Putting $N=1$ in Eqs.~(\ref{green_mul}), (\ref{xx_mul})
and (\ref{pp_mul}), we have
\begin{align}
&\langle \hat{x}^2 \rangle_{\rm mfG}=\frac{1}{M\beta}\sum_{r=-\infty}^\infty \frac{1}{\nu_r^2+\Omega^2+\zeta_r}, \label{xx_n1}\\
&\langle \hat{p}^2 \rangle_{\rm mfG}=\frac{M}{\beta}\sum_{r=-\infty}^\infty \frac{\Omega^2+\zeta_r}{\nu_r^2+\Omega^2+\zeta_r},
\end{align}
Taking the coupling strength $\gamma$ in $\zeta_r$ to infinity (see Eq.~(\ref{ohmic})), 
we see that the infinite sum in $\langle \hat{p}^2\rangle_{\rm mfG}$ diverges. We also find that 
only the $r=0$ term in Eq.~(\ref{xx_n1}) survives, which gives $\langle \hat{x}^2 \rangle_{\rm mfG}\to 1/(M\beta\Omega^2)$ as
$\gamma\to \infty$. As can be seen from Eqs.~(\ref{mfg_element_mul}),
the diverging momentum covariance indicates that the matrix element $\langle x \vert \rho_{\rm mfG} \vert x^\prime\rangle$ is
proportional to $\delta(x-x^\prime)$ i.e.\ diagonal in 
eigenstates of $\hat{x}$ in the USC limit. This point was recognized recently in Ref.~\cite{kumar2024ultrastrong}.
The actual value of the position covariance in the USC limit indicates that $\rho_{\rm USC}$ is given by
the Gibbs state of
an effective Hamiltonian of the form $\hat{H}_{\rm USC}=(1/2)M\Omega^2 \hat{x}^2$.
This is exactly the system Hamiltonian projected onto the position eigenstate as done in Eq.~(\ref{usc}). 
It is, however, different from another form of the USC limit suggested in Refs.~\cite{Goyal_Kawai, Orman_Kawai},
which is equal to $\sum_n \hat{P}_n \hat{\rho}_{\rm G} \hat{P}_n=\int dx \vert x\rangle\langle x \vert
\hat{\rho}_{\rm G} \vert x\rangle\langle x \vert$ in our notation. This expression would give the position covariance 
as $(\hbar/(2M\Omega))\coth (\beta\hbar\Omega/2)$. This can be seen by using Eq.~(\ref{summation}) on the diagonal element
in Eq.~(\ref{mfg_element_mul}) with $G^{(0)}$ replacing $G$ since we are using $\hat{\rho}_{\rm G}$. 
But this is clearly in disagreement with the actual 
USC limit $1/(M\beta\Omega^2)$ obtained from the exact expression Eq.~(\ref{xx_n1}) except for very high temperatures.

The second point why we study the USC limit in our model is the following. 
We note that the USC limit suggested in Eq.~(\ref{usc}) is characterized by the effective Hamiltonian which is a
diagonal projection onto the eigenstates of the system operator that couples to the bath. In our model,
this system operator is only local as only a part of the system couples to the bath. 
Therefore, using our exact result, we can check whether the USC limit in Eq.~(\ref{usc}) still holds 
for the case where the coupling operator is local in space. To do this 
we focus on the simplest case $N=2$, where the positions of the oscillator are $\hat{x}_0$ and $\hat{x}_1$ and 
only $\hat{x}_0$ is coupled to the bath. 
The normal mode frequencies from Eq.~(\ref{evalue}) are
$\Omega^2_{0}=\Omega^2 - 2\lambda$ and $\Omega^2_{1}=\Omega^2 + 2\lambda$, and
from Eq.~(\ref{G0}), we have
\begin{align}
&G^{(0)}_{00}(z)=G^{(0)}_{11}(z)=\frac 1 {2} \left[ 
\frac 1 {z^2+\Omega^2_0} +\frac{1}{z^2+\Omega^2_{1}}\right] , \label{N2G00}\\
&G^{(0)}_{01}(z)=G^{(0)}_{10}(z)=\frac 1 {2} \left[ 
\frac 1 {z^2+\Omega^2_0} - \frac{1}{z^2+\Omega^2_{1}}\right] .
\end{align}

The USC limit of the two-oscillator system is obtained by  taking $\gamma\to\infty$ in 
Eqs.~(\ref{xx_full}) and (\ref{pp_full}). We have 
\begin{align}
&\langle \hat{x}^2_0 \rangle_{\rm mfG}\xrightarrow[\gamma\to\infty]{} \frac{G^{(0)}_{00}(0)}{M\beta}
=\frac 1{2M\beta}\left(\frac 1 {\Omega^2_0}+\frac 1 {\Omega^2_1}\right) ,  \label{x02_usc} \\
&\langle \hat{x}_0\hat{x}_1 \rangle_{\rm mfG}\xrightarrow[\gamma\to\infty]{} \frac{G^{(0)}_{01}(0)}{M\beta}
=\frac 1{2M\beta}\left(\frac 1 {\Omega^2_0}-\frac 1 {\Omega^2_1}\right), \label{x01_usc}
\end{align}
and
\begin{align}
\langle \hat{x}^2_1 \rangle_{\rm mfG} & \xrightarrow[\gamma\to\infty]{} \frac{G^{(0)}_{11}(0)}{M\beta}
+\frac{2}{M\beta}\sum_{r=1}^\infty \frac{ G^{(0)}_{00}(\nu_r)^2-G^{(0)}_{01}(\nu_r)^2}{ G^{(0)}_{00}(\nu_r)} \nonumber \\
&=\frac 1{2M\beta}\left(\frac 1 {\Omega^2_0}+\frac 1 {\Omega^2_1}\right)
+\frac{2}{M\beta}\sum_{r=1}^\infty \frac{1}{\nu^2_r+\Omega^2}. \label{x12_usc}
\end{align}
For the momentum covariances, we can easily see from Eq.~(\ref{pp_full}) that
$\langle \hat{p}^2_0 \rangle_{\rm mfG}\to \infty$ and $\langle \hat{p}_0\hat{p}_1 \rangle_{\rm mfG}\to 0$
as $\gamma \to\infty$ and
\begin{align}
\langle \hat{p}^2_1  \rangle&_{\rm mfG}  \xrightarrow[\gamma\to\infty]{}
\frac{M}{\beta}\sum_{r=-\infty}^{\infty}\left[ 1-\nu^2_r \frac{ G^{(0)}_{00}(\nu_r)^2-G^{(0)}_{01}(\nu_r)^2}{ G^{(0)}_{00}(\nu_r)} \right] \nonumber\\
&= \frac{M}{\beta}\sum_{r=-\infty}^{\infty} \frac{\Omega^2}{\nu^2_r+\Omega^2}=\frac{M\hbar\Omega}{2}\coth\left(\frac{\beta\hbar\Omega}{2}\right).
\label{p12_usc}
\end{align}

We now check whether the USC limit,
Eq.~(\ref{usc}) conjectured in Ref.~\cite{cresser_weak_2021} is consistent with this large-$\gamma$ limit obtained from
the exact results.
In the present two-oscillator case, the USC limit of Ref.~\cite{cresser_weak_2021} is characterized by the effective Hamiltonian 
which is a diagonal projection of $\hat{H}_{\rm S}$ onto the eigenstates of $\hat{x}_0$. As we have seen in 
the one-oscillator case, it amounts to suppressing in the system Hamiltonian the kinetic energy ($\hat{p}^2_0/(2M)$) of the oscillator
in contact with the heat bath. The matrix element of $\hat{\rho}_{\rm USC}$ in Eq.~(\ref{usc}) is then given by
\begin{align}
& \langle \bar{x}_0, \bar{x}_1 \vert
 \hat{\rho}_{\rm USC} \vert \bar{x}_0, \bar{x}^\prime_1 \rangle  \label{rho_USC}\\
=& \frac 1{Z_{\rm USC}}\int_{x_1(0)=\bar{x}_1^\prime}^{x_1(\beta\hbar)=\bar{x}_1} 
\mathcal{D}x_1 (\tau) 
  \exp\left [-\frac 1 {\hbar} S_{\rm USC}[x_1;\bar{x}_0] \right], \nonumber 
\end{align}
where the system Euclidian action is given by
 \begin{align}
 S_{\rm USC}[x_1;& \bar{x}_0]=\int_0^{\beta\hbar} d\tau \; \Big[ \frac M 2 \left( \frac{d x_1 (\tau)} {d\tau}\right)^2 \nonumber \\
& +\frac M 2 \Omega^2 \left( \bar{x}^2_0+ x^2_1(\tau)   \right) -2M\lambda \bar{x}_0 x_1(\tau) \Big]  \label{S_USC}
\end{align} 
with
$Z_{\rm USC}=\int d\bar{x}_0 d\bar{x}_1  \langle \bar{x}_0, \bar{x}_1 \vert
 \hat{\rho}_{\rm USC} \vert \bar{x}_0, \bar{x}_1 \rangle$.
 We can evaluate this path integral by seeking the stationary path solution
 as done in Sec.~\ref{sec:setup} and in Appendices \ref{app:2} and \ref{app:3}. 
 Detailed calculations are given in Appendix \ref{app:5}. We obtain
 \begin{align}
 &\langle \bar{x}_0, \bar{x}_1 \vert
 \hat{\rho}_{\rm USC} \vert \bar{x}_0, \bar{x}^\prime_1 \rangle = \tilde{C}\exp\Bigg[ 
- \frac M {2\hbar} \Big\{A (\bar{x}_1^\prime-\bar{x}_1)^2  \nonumber \\
&+\frac 1 F \left(\frac{\bar{x}_1+\bar{x}_1^\prime}{2}- \frac{2\lambda}{\Omega^2}\bar{x}_0\right)^2 
 +\beta\hbar\Omega^2\left(1-\frac{4\lambda^2}{\Omega^4}\right)\bar{x}^2_0\Big\}\Bigg],
 \label{matrix_usc}
 \end{align}
 where
 \begin{align}
 &A=\frac{1}{\beta\hbar}\sum_{r=-\infty}^\infty \frac {\Omega^2}{\nu^2_r+\Omega^2}=\frac {\Omega} {2}\coth \left( \frac{\beta\hbar\Omega}{2} \right) , \\
 &F=\frac{1}{\beta\hbar}\sum_{r=-\infty}^\infty \frac 1{\nu^2_r+\Omega^2}=\frac {1} {2\Omega}\coth \left( \frac{\beta\hbar\Omega}{2} \right),
 \end{align}
and $\tilde{C}$ is determined from
$1=\int d\bar{x}_0 d\bar{x}_1 \;  \langle \bar{x}_0, \bar{x}_1 \vert
 \hat{\rho}_{\rm USC} \vert \bar{x}_0, \bar{x}_1 \rangle$.
 
 The covariances are now calculated by evaluating Gaussian integrals such as
 \begin{align}
 \langle \hat{x}^2_0\rangle_{\rm USC} & =\int d\bar{x}_0 d\bar{x}_1 \;  \bar{x}_0^2
  \langle \bar{x}_0, \bar{x}_1 \vert
 \hat{\rho}_{\rm USC} \vert \bar{x}_0, \bar{x}_1 \rangle \nonumber \\
 &=\frac 1 {M\beta} \frac{\Omega^2}{\Omega^4-4\lambda^2},
 \end{align}
 and 
 \begin{align}
 \langle \hat{x}_0 \hat{x}_1\rangle_{\rm USC} & =\int d\bar{x}_0 d\bar{x}_1  \; \bar{x}_0\bar{x}_1
  \langle \bar{x}_0, \bar{x}_1 \vert
 \hat{\rho}_{\rm USC} \vert \bar{x}_0, \bar{x}_1 \rangle  \nonumber \\
 &=\frac{2\lambda}{\Omega^2}  \langle \hat{x}^2_0\rangle_{\rm USC} =
 \frac 1 {M\beta} \frac{2\lambda}{\Omega^4-4\lambda^2},
 \end{align}
 where we have shifted the integration variable $\bar{x}_1\to \bar{x}_1+2\lambda \bar{x}_0/\Omega^2$
 in the second integral.
 Note that these are exactly the same as Eqs.~(\ref{x02_usc}) and (\ref{x01_usc}), respectively.
 The remaining part is given by
 \begin{align}
 \langle \hat{x}^2_1\rangle_{\rm USC} & =\int d\bar{x}_0 d\bar{x}_1  \; \bar{x}^2_1
  \langle \bar{x}_0, \bar{x}_1 \vert
 \hat{\rho}_{\rm USC} \vert \bar{x}_0, \bar{x}_1 \rangle  \nonumber \\
 &= \frac{\hbar}{M} F + \frac{4\lambda^2}{\Omega^4}  \langle \hat{x}^2_0\rangle_{\rm USC}
 \end{align}
 We can easily verify that this coincides with Eq.~(\ref{x12_usc}).
 For the momentum covariance, we can see that only the off-diagonal part ($\exp[-MA(\bar{x}_1^\prime-\bar{x}_1)^2/(2\hbar)]$) is relevant.
Therefore, we have
\begin{align}
\langle \hat{p}^2_1\rangle_{\rm USC} =M\hbar A,
\end{align}
which is just Eq.~(\ref{p12_usc}). In summary, we have verified that the USC limit
given as Eq.~(\ref{usc}) is still valid even if the projection is onto a \textit{local} system operator.

\begin{figure}
	\resizebox{0.75\columnwidth}{!}{\includegraphics{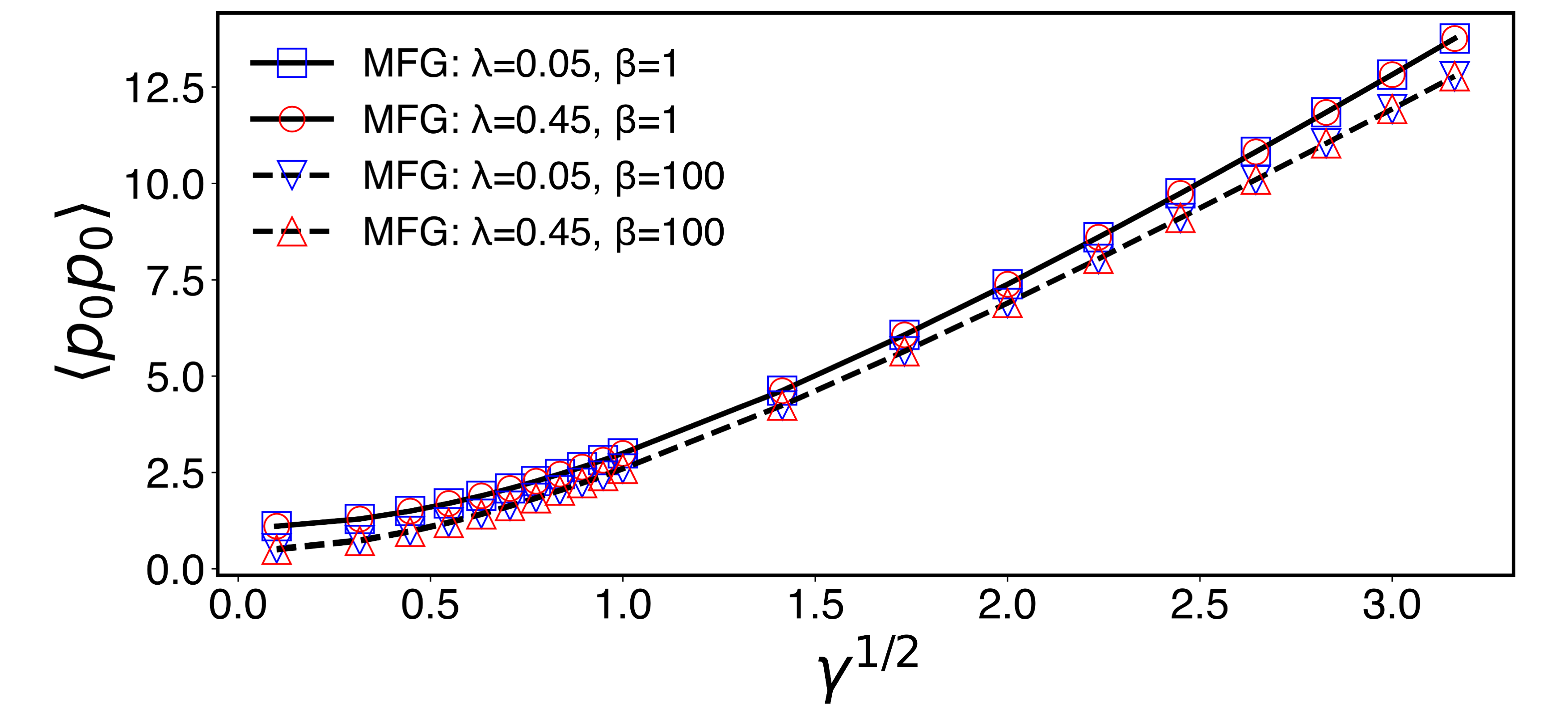}}
	\caption{ The momentum covariance $\langle\hat{p}^2_0\rangle_{\rm mfG}$ with respect to the MFG state at the point of contact with the heat bath as a function of $\gamma^{1/2}$ for $N=20$ coupled oscillators.  For various temperatures and couplings between the oscillators, it increases as $\gamma^{1/2}$ in the large-$\gamma$ limit. 
	}
	\label{fig:large_g}
\end{figure}

As we have shown, the USC limit is characterized by the diverging momentum covariance $\langle \hat{p}^2_0\rangle_{\rm mfG}$
at the point of contact with the heat bath. This is related to the projection of the effective Hamiltonian 
on the space of system local coupling operator $\hat{x}_0$ in the USC limit. The divergence of $\langle \hat{p}^2_0\rangle_{\rm mfG}$
as $\gamma\to\infty$ can be made more quantitative as follows.
For the $N=2$ case with the Ohmic bath given in Eq.~(\ref{ohmic}), we can actually show that as $\gamma\to\infty$
\begin{align}
\langle \hat{p}^2_0 \rangle_{\rm mfG} \sim M\hbar\sqrt{\frac{\omega_D\gamma }{2}}.
\end{align} 
The derivation is given in Appendix \ref{app:6}. In fact we find that this large-$\gamma$ behavior of $\sim\gamma^{\frac 1 2}$
continues to be true even for general number of oscillators.
Figure \ref{fig:large_g} shows $\langle \hat{p}^2_0 \rangle_{\rm mfG}$ as a function of $\gamma$ for $N=20$ oscillators for various temperatures 
and inter-oscillator coupling constants. For large $\gamma$, it increases with $\gamma^{\frac 1 2}$. Therefore, we conclude that
in the USC limit $\langle \hat{p}^2_0 \rangle_{\rm mfG}$ goes to infinity, which results in the projection of the MFG state
onto the eigenstate of $\hat{x}_0$ that couples to the heat bath.

\section{System in contact with multiple heat baths at the same temperature}
\label{sec:multiple}

In this section, we apply the general formalism developed above to the case where 
there are multiple heat baths at the same inverse temperature $\beta$. We first focus on the case where 
there are two baths and 
the system makes contact with them at the locations given by $\mathcal{B}=\{\alpha,\alpha^\prime\}$.
We then generalize to the multiple-bath case.
For the two-bath case, the diagonal matrix $\bm{\Sigma}$ in Eq.~(\ref{green_mul}) has only two nonzero elements, namely
$\Sigma_{\alpha \alpha}(\nu_r)=\zeta^{(\alpha)}_r$ and $\Sigma_{\alpha^\prime\alpha^\prime}(\nu_r)=\zeta^{(\alpha^\prime)}_r$.
For later use, we denote by $\bm{\bar{\Sigma}}$ the $2\times 2$ matrix composed of these nonzero elements,
which belongs to the subspace spanned by $\{\alpha,\alpha^\prime\}$.

As in the case of the single bath, we try to express the Green's
function in Eq.~(\ref{green_mul}) in terms of $\bm{G}^{(0)}$ which is obtained
in the absence of the coupling to the bath and the correction to it.
Now from Eq.~(\ref{green_exp}), we can rewrite this as
\begin{align}
\bm{G}(\nu_r)=\bm{G}^{(0)}(\nu_r) - \bm{G}^{(0)}(\nu_r) \bm{\Gamma}(\nu_r) \bm{G}^{(0)}(\nu_r),
\label{green2}
\end{align}
where
\begin{align}
\bm{\Gamma}(\nu_r)=&\bm{\Sigma}- \bm{\Sigma} \bm{G}^{(0)}   \bm{\Sigma} +\bm{\Sigma} \bm{G}^{(0)}   \bm{\Sigma}  \bm{G}^{(0)}   \bm{\Sigma} -\cdots .
\label{green_exp2}
\end{align}
From the structure of these terms, we realize that all the elements of $\Gamma_{ij}$ are zero except when
$i$ and $j$ are equal to $\alpha$ or $\alpha^\prime$.  We again denote this nonzero  $2\times 2$ matrix 
by $\bm{\bar{\Gamma}}$. We note that Eq.~(\ref{green2}) has the same structure as Eq.~(\ref{deltag1})
for the one-bath case. The only difference is that $\bm{\Gamma}$ now contains a nonzero $2\times 2$ sub-matrix
instead of the single nonvanishing component. 
In order to calculate this $\bm{\bar{\Gamma}}$, 
we notice that Eq.~(\ref{green_exp2}) can be considered as a relation among
$2\times 2$ matrices $\bm{\bar{\Sigma}}$, $\bm{\bar{\Gamma}}$ and 
$\bm{\bar{G}}^{(0)}$, the last of which is a matrix composed of $G^{(0)}_{ij}$
with $i,j \in \{\alpha,\alpha^\prime\}$.
By summing the infinite series in Eq.~(\ref{green_exp2}), we can write 
\begin{align}
\bm{\bar{\Gamma}}(\nu_r)=& (\bm{1}+ \bm{\bar{\Sigma}} \bm{\bar{G}}^{(0)} )^{-1} \bm{\bar{\Sigma}} \nonumber \\
=&\left[  \bm{\bar{\Sigma}}^{-1} +  \bm{\bar{G}}^{(0)} \right]^{-1}.
\label{kernel}
\end{align}
So $\bm{\bar{\Gamma}}(\nu_r)$ is obtained by simply inverting the known
$2\times2$ matrix. The correction term $\Delta\bm{G}$ can then be calculated using Eq.~(\ref{deltag1})
with newly obtained $\bm{\Gamma}$. 

This procedure can easily be generalized to a system in contact with $m$ reservoirs with $m>2$. 
All one has to do is to calculate the $m\times m$ kernel matrix $\bm{\bar{\Gamma}}$ from Eq.~(\ref{kernel})
and calculate the Green's function for the MFG state from Eq.~(\ref{green2}).

\subsection{Example: Chain of Harmonic Oscillators in Contact with Two Heat Baths at the Same Temperature}

Here we apply this method to the ring of oscillators studied in Sec.~\ref{sec:chain1}. We focus on the case where there
are even number of oscillators and the two baths are connected to the oscillators at $x_0$ and $x_{\frac N 2}$.
In the present notation, we take $\alpha=0$ and $\alpha^\prime=\frac N 2$. See Fig.~\ref{fig:osc} (b) for a schematic diagram
describing the situation.
We assume that the two baths are
described by the Ohmic spectral densities.
In particular, we assume that the baths at $\alpha=0$ and $\alpha^\prime=\frac N 2$ are described by $\zeta_r$
and $\zeta^\prime_r$, respectively, which are given by
\begin{align}
\zeta_r =2\gamma_1 \frac{\omega_D \vert \nu_r \vert }{\vert \nu_r \vert+\omega_D},~~~
\zeta^\prime_r =2\gamma_2 \frac{\omega_D \vert \nu_r \vert }{\vert \nu_r \vert+\omega_D}. \label{ohmic2}
\end{align}
with in general two distinct coupling strengths $\gamma_1$ and $\gamma_2$.

In order to calculate $\bm{\bar{\Gamma}}$ in Eq.~(\ref{kernel}), we first note that in the subspace of $\{0,\frac N 2\}$  
\begin{align}
\label{sigmabar}
\bm{\bar{\Sigma}}(\nu_r)=
\begin{pmatrix}
\zeta_r & 0  \\
0 & \zeta^\prime_r 
\end{pmatrix}.
\end{align}
It is then straightforward to evaluate $\bm{\bar{\Gamma}}(\nu_r)$ from Eq.~(\ref{kernel}). 
From Eqs.~(\ref{kernel}) and (\ref{sigmabar}), we have
\begin{align}
\bm{\bar{\Gamma}}(\nu_r)=
\begin{pmatrix}
G^{(0)}_{00}+\frac 1 {\zeta_r} & G^{(0)}_{0\frac{N}{2}} \\
G^{(0)}_{\frac{N}{2}0}  & G^{(0)}_{\frac{N}{2}\frac{N}{2}}+\frac 1 {\zeta^{\prime }_r} 
\end{pmatrix}^{-1}.
\end{align}
From Eq.~(\ref{G0}), we can see that only two of the components  of $\bm{G}^{(0)}$ are independent, which we denote by
\begin{align}
&G_+ (\nu_r)\equiv G^{(0)}_{00}(\nu_r) =  G^{(0)}_{\frac{N}{2}\frac{N}{2}} (\nu_r),\\
&G_- (\nu_r) \equiv  G^{(0)}_{0\frac{N}{2}}(\nu_r) =G^{(0)}_{\frac{N}{2}0} (\nu_r) .
\end{align}
By inverting the above $2\times 2$ matrix, we obtain
\begin{align}
&\bar{\Gamma}_{00}(\nu_r) =\frac {\zeta_r} {D(\nu_r)} (1+\zeta^\prime_r G_+(\nu_r)), \\
&\bar{\Gamma}_{\frac{N}{2}\frac{N}{2}} (\nu_r) = \frac {\zeta^\prime_r} {D(\nu_r)} (1+\zeta_r G_+(\nu_r)),
\end{align}
and
\begin{align}
\bar{\Gamma}_{0\frac{N}{2}}(\nu_r)=\bar{\Gamma}_{\frac{N}{2}0}(\nu_r)=-\frac{\zeta_r\zeta^\prime_r}{D(\nu_r)}G_-(\nu_r),
\end{align}
where
\begin{align}
D(\nu_r) =& (1+\zeta_r G_+(\nu_r))(1+\zeta^\prime_r G_+(\nu_r)) \nonumber \\
&-\zeta_r\zeta^\prime_rG^2_-(\nu_r).
\end{align}
Using these results and Eq.~(\ref{deltag1}), we can write the correction to the system Gibbs state as
\begin{align}
\Delta G_{ij}(\nu_r) = - & \sum_{\alpha=0,N/2} \sum_{\alpha^\prime=0,N/2}  G^{(0)}_{i\alpha}(\nu_r) \nonumber \\
&\times \Gamma_{\alpha\alpha^\prime} (\nu_r) G^{(0)}_{\alpha^\prime j}(\nu_r).
\end{align}

\begin{figure}
	\resizebox{0.95\columnwidth}{!}{\includegraphics{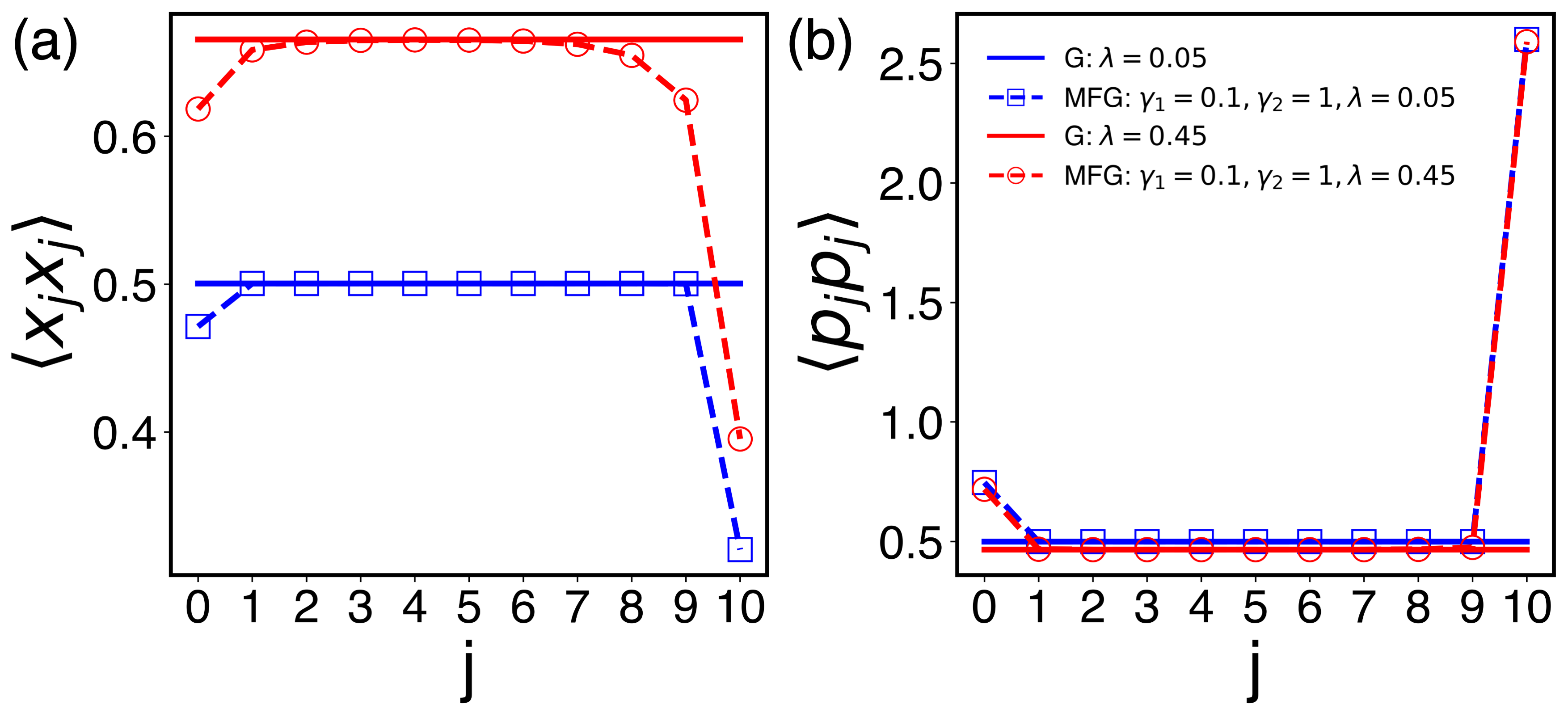}}
	\caption{The covariances (a) $\langle\hat{x}^2_j\rangle$ and (b) $\langle\hat{p}^2_j\rangle$ as functions of the position $j$ of the $N=20$ oscillators.
	As in Fig.~\ref{fig:b1beta100}, the solid horizontal lines are obtained using the system Gibbs (G) state. 
	The system is in contact with two heat baths maintained at the same inverse temperature $\beta=100$ at $j=0$ and $j=10$ with coupling strengths 
	$\gamma_1=0.1$ and $\gamma_2=1$, respectively. The data points and the dashed lines are obtained using the MFG state. 
	The covariances for two different values of the inter-oscillator coupling constants $\lambda=0.05$ (squares) and $\lambda=0.45$ (circles)
	are displayed.
	}
	\label{fig:b2beta100}
\end{figure}

We now calculate the covariances using Eqs.~(\ref{deltaxx}) and (\ref{deltapp}) by evaluating the infinite sums numerically.
In Fig.~\ref{fig:b2beta100}, we show results of these calculations for a system coupled to two heat baths maintained at some low temperature $\beta=100$
with different coupling strengths. As in the one-bath case, the MFG state shows a skin effect for each heat bath, where the effect of the coupling to the bath
quickly decreases as one moves away from the contact points. For a relatively large system size $N=20$, Fig.~\ref{fig:b2beta100} shows 
a separate skin effect for each heat bath.  

\section{Discussion and Summary}

We have presented a detailed procedure of obtaining the exact quantum MFG state of a coupled harmonic system
interacting with multiple heat baths at the same temperature. We have provided a nonperturbative method 
to calculate the covariances with respect to the MFG state. For the specific examples of 
the chain of harmonic oscillators, we have calculated these covariances as we vary the physical parameters such as the 
temperature, the coupling strength to the bath and the inter-oscillator coupling constant and compare them with those
calculated with respect to the system Gibbs state. We found that there is a skin effect similar to 
that found in Ref.~\cite{burke_structure_2024}, where the effect of the coupling
to the bath quickly disappears as a function of the distance of the system-bath boundary. We have also investigated 
quantitatively the skin depth and found that it is quite short-ranged and insensitive to the coupling strength to the bath.
Using these exact results, we were also able to confirm that the ultrastrong coupling limit of the MFG state found recently in Ref.~\cite{cresser_weak_2021}
is valid even in the case where the system coupling operator is local and has a continuous spectrum.

An interesting aspect of the MFG state which has not been explored in this paper is its relation to the thermalization of an open quantum system.
If we consider the dynamics of an open quantum system coupled with a heat bath,  we expect the system reaches the MFG state as a
steady state. The dynamics is often described 
in terms of quantum master equations, most of which are based on weak coupling approximations \cite{OQS_Breuer}.
The Lindblad equation \cite{Lindblad,GKS} has been studied extensively as it guarantees the complete positivity.
The standard derivation of the Lindblad equation
involves a series of weak coupling approximations, which 
forces the local coupling system operator to be written
in terms of eigenoperators with respect to the whole system Hamiltonian  \cite{OQS_Breuer}
and in turn makes it global. The global Lindblad equation is often referred as a ultraweak coupling theory \cite{trushechkin_open_2022}
as its steady state is given by the system Gibbs state, which is the zero coupling limit
of the MFG state. One can keep the local nature of the coupling system operator in the Lindblad
equation \cite{landi_nonequilibrium_2022}. The local Lindblad equation is shown to be consistent
with local conservation laws \cite{tupkary_2023} unlike the global one, but the steady state 
is no longer given by the Gibbs or MFG state \cite{tupkary_fundamental_2022,carlen2024}.
If we relax the condition of complete positivity, one has the Redfield equation, \cite{Redfield} which 
can again be derived from the weak coupling approximations
corresponding to the second-order perturbative expansion in the system-bath coupling.
Now, the steady state of the Redfield equation is known to be equal to the perturbative MFG state valid up to the same order 
of perturbation theory \cite{thingna_generalized_2012, tupkary_fundamental_2022, Lee_2022}. 
However, as in the global Lindblad equation, the coupling system operator becomes global.
This seems to be in contrast to our finding here that the local nature of the system operator playing a role in the structure 
of the MFG state. It would be interesting to see how the structure of the MFG state found in this paper
manifest in the steady states of various weak-coupling quantum master equations
for an extended system.

In this paper, we have obtained using the path integral method 
the equilibrium MFG state of a system of harmonic oscillators in contact with multiple heat baths 
at the same temperature. It would be interesting to see if one can apply the similar path integral formalism to study
steady states of a nonequilibrium system interacting with heat baths 
maintained at different temperatures.  The path integral formalism has been used 
in such nonequilibrium situation for different physical contexts \cite{martinez_dynamics_2013, estrada_quantum}. 
Unlike the present equilibrium problem, however,
one would have to keep track of the full time dependence of the nonequilibrium system to investigate the steady states in the long time limit.
This is left for future study.

\begin{acknowledgments}
This work was supported by NRF
grant funded by the Korea government (MSIT) (RS-2023-00276248).
\end{acknowledgments}

\appendix
\section{The Influence Functional}
\label{app:1}

We first write $K_\alpha(\tau)$ in Eq.~(\ref{Ka}) which is periodic in the interval $[0,\beta\hbar]$ as a Fourier series
\begin{align}
K_\alpha(\tau) =\frac 1 {\beta\hbar}\sum_{r=-\infty}^{\infty} g^{(\alpha)}_r e^{i\nu_r \tau}
\label{Ka_exp}
\end{align}
with $\nu_r=2\pi r/(\beta\hbar)$. Then using Eq.~(\ref{Ka}) we have
\begin{align}
g^{(\alpha)}_r &=\int_0^{\beta\hbar} d\tau\; e^{-i\nu_r \tau}K_\alpha (\tau)=\mu_\alpha - M\zeta^{(\alpha)}_r,
\end{align}
where $\mu^{(\alpha)}$ and $\zeta^{(\alpha)}_r$ are defined in Eq.~(\ref{mua}) and Eq.~(\ref{zetara}), respectively.
Putting this back into Eq.~(\ref{Ka_exp}) and using the Poisson sum rule, we have
\begin{align}
K_\alpha(\tau)=\mu_\alpha\sum_{r=-\infty}^\infty \delta(\tau- r\beta\hbar)-M L_\alpha(\tau),
\end{align}
where $L_\alpha(\tau)$ is given in Eq.~(\ref{La}).
We therefore have
\begin{align}
&\int_0^{\beta\hbar} d\tau \int_0^\tau d\tau'\; K_\alpha(\tau-\tau')x_\alpha(\tau)x_\alpha(\tau')
=   \frac{\mu_\alpha}{2}\int_0^{\beta\hbar}d\tau\; x_\alpha^2(\tau)  \nonumber  \\
&-\frac M 2\int_0^{\beta\hbar} d\tau \int_0^{\beta\hbar} d\tau'\; L_\alpha(\tau-\tau')x_\alpha(\tau)x_\alpha(\tau').
\end{align}

\section{Stationary Path Solution}
\label{app:2}

We follow closely the procedure described in Ref.~\cite{grabert_quantum_1988}.
We solve the stationary path condition Eq.~(\ref{stationary_mul}) using the Fourier series given in
Eq.~(\ref{xexpand_mul}), which is defined outside the interval $[0,\beta\hbar]$ as well. 
The derivatives of $\bm{x}(\tau)$
with respect to $\tau$ then produce singularities at the endpoints $\tau=r\beta\hbar$, $r=0,\pm 1,\pm2, \cdots$.
In order to account for these singularities, we rewrite Eq.~(\ref{stationary_mul}) as
\begin{align}
&- \frac{d^2 x_j(\tau)}{d\tau^2}+\frac 1 M \sum_{k=0}^{N-1}\Lambda_{jk}x_k(\tau) \label{stationary_delta}\\
&+\sum_{\alpha\in\mathcal{B}}\int_0^{\beta\hbar}d\tau'\; L_\alpha(\tau-\tau')x_\alpha(\tau') 
= - a_j \tilde{\delta}'(\tau) - b_j \tilde{\delta}(\tau), \nonumber 
\end{align}
for some constants $a_i$ and $b_i$ to be determined later,
where 
\begin{align}
\tilde{\delta}(\tau)\equiv\sum_{n=-\infty}^\infty \delta(\tau-n\beta\hbar)=\frac 1 {\beta\hbar}\sum_{r=-\infty}^\infty e^{i\nu_r \tau}
\end{align}
is the discrete delta function.
In terms of the Fourier mode, we have for $j=0,1,\cdots,N-1$,
\begin{align}
&  \nu^2_r x_{j,r} +\frac 1 M \sum_k \Lambda_{jk} x_{k,r}+\sum_{\alpha\in \mathcal{B}}\zeta^{(\alpha)}_r x_{\alpha,r} \nonumber \\
 =&-i\nu_r a_j  - b_j .
 \label{xr_sol_app}
\end{align}
This equation is to be solved with the boundary condition $\bm{x}(0^+)=\bm{\bar{x}'}$ and 
$\bm{x}(\beta\hbar)=\bm{x}(0^-)=\bm{\bar{x}}$. The solution can be written as
\begin{align}
x_{j,r}=- \sum_{k=0}^{N-1}G_{jk}(\nu_r)\left[ i\nu_r a_k + b_k\right],
\label{xrsol}
\end{align}
where
\begin{align}
[\bm{G}^{-1}(\nu_r)]_{jk}=\nu^2_r\delta_{jk}+\frac 1 M \Lambda_{jk}+\Sigma_{jk}(\nu_r)
\end{align}
and $\Sigma_{jk}(\nu_r)$ is a diagonal matrix with non-vanishing elements $\zeta^{(\alpha)}_r$ only for $j=k=\alpha$, that is
\begin{align}
\Sigma_{jk}(\nu_r)=\delta_{jk}\sum_{\alpha\in\mathcal{B}}\delta_{j\alpha}\zeta^{(\alpha)}_r.
\label{Gammajk}
\end{align} 
The discontinuity of $x_j(\tau)$ at $\tau=0$ is responsible for term 
proportional to $\tilde{\delta}'(\tau)$. In fact we can easily see that
\begin{align}
a_j= \bar{x}^\prime_j -\bar{x}_j.
\label{ajsol}
\end{align}
The discontinuity at $\tau=0$ can also be read off from Eq.~(\ref{xrsol}) by forming 
$
x_j(\tau)=\sum_r x_{j,r}\exp(i\nu_r \tau)
$
and taking the $\tau\to 0$ limit. We first note that
for large $\nu_r$, $G_{jk}(\nu_r)\sim 1/\nu^2_r$. Therefore
the term proportional to $a_k$ in Eq.~(\ref{xrsol}) contains a contribution which 
goes as $1/\nu_r$ for large $r$. This kind of contribution when summed over all nonzero $r$ 
signals a discontinuity at $\tau=0$. More explicitly, we write a part of $x_j(\tau)$ 
obtained from the first 
term on the right hand side of Eq.~(\ref{xrsol}) as
\begin{align}
\frac{1}{\beta\hbar} \sum_{r=-\infty}^{\infty\prime} \sum_k\frac{1}{i\nu_r} [ \mathcal{G}_{jk}+ (\nu^2_r G_{jk}(\nu_r)-\mathcal{G}_{jk})]a_k
e^{i\nu_r\tau},
\label{aipart}
\end{align}
where $\mathcal{G}_{jk}=\lim_{z\to \infty}z^2 G_{jk}(z)$ and the prime indicates that the absence of the $r=0$ term.
When taking the $\tau\to 0$ limit, the term in the round parenthesis  vanishes as $G_{jk}(\nu_r)$ is even in $\nu_r$.
We also note that a sawtooth wave  $f(\tau)=1/2-\tau/(\beta\hbar)$ defined in $[0,\beta\hbar]$ has a Fourier expansion
\begin{align}
f(\tau)=\frac 1 {\beta\hbar}
\sum_{r=-\infty}^{\infty\prime} \frac 1 {i\nu_r}e^{i\nu_r \tau}.
\end{align} 
Note that the sawtooth wave has a discontinuity as $f(0^+)=1/2$ and $f(0^-)=-1/2$.
Therefore, Eq.~(\ref{aipart}) becomes $(1/2)\sum_k\mathcal{G}_{jk}a_k$ as $\tau\to 0^+$
and  $-(1/2)\sum_k\mathcal{G}_{jk}a_k$ as $\tau\to 0^-$.
Finally by taking the $\tau=0^{\pm}$ limit of $x_j(\tau)$, we have from Eq.~(\ref{xrsol})
\begin{align}
&x_j(0^+) =   \sum_k\left[ \frac 1 2 \mathcal{G}_{jk}a_k - F_{jk} b_k\right]=\bar{x}'_j  \label{eq_b1}\\
&x_j(0^-) =   \sum_k\left[ -\frac 1 2 \mathcal{G}_{jk}a_k - F_{jk} b_k\right]=\bar{x}_j, \label{eq_b2}
\end{align}
where 
\begin{align}
F_{jk}\equiv \frac 1 {\beta\hbar}\sum_{r=-\infty}^{\infty} G_{jk}(\nu_r).
\end{align}
Adding these two equations, we have
\begin{align}
b_j =-\frac 1 {2} \sum_{k=0}^{N-1}(\bm{F}^{-1})_{jk}\left ( \bar{x}_k +\bar{x}'_k\right).
\label{bjsol}
\end{align}
Inserting Eqs.~(\ref{bjsol}) and (\ref{ajsol}) into Eq.~(\ref{xrsol}), we have Eq.~(\ref{xr_mul}).

\section{Evaluation of the MFG State from the Stationary Solution}
\label{app:3}

We write the actions inside the exponential of Eq.~(\ref{mfg_path_mul}) using the stationary solution Eq.~(\ref{xr_mul}) with
Eq.~(\ref{xrsol}).
We note that the derivative $dx_j/d\tau$ expressed in terms of the Fourier modes contains the delta function
singularities coming from the discontinuities. By removing these singularities and using the matrix notations
$\bm{a}$ for Eq.~(\ref{ajsol}) and $\bm{b}$ for Eq.~(\ref{bjsol}), we can write
\begin{align}
&\frac{d \bm{x}(\tau)}{d\tau}=\frac 1{\beta\hbar}\sum_{r=-\infty}^\infty i\nu_r \bm{x}_{r}e^{i\nu_r \tau}-\bm{a} \tilde{\delta}(\tau)  
\label{dxdtau} \\
=&-\frac 1{\beta\hbar}\sum_{r=-\infty}^\infty \Big[\left\{ \bm{1}-\nu^2_r \bm{G}(\nu_r)\right\}\bm{a} 
+ i\nu_r  \bm{G}(\nu_r)\bm{b})\Big] e^{i\nu_r \tau}. \nonumber
\end{align}
\begin{widetext}
Then we have
\begin{align}
\frac M 2\int_0^{\beta\hbar}d\tau\; \frac{d \bm{x}^{\rm T}(\tau)}{d\tau}\frac{d \bm{x}(\tau)}{d\tau} =\frac M 2
\frac 1{\beta\hbar}\sum_{r=-\infty}^\infty &\Big[ \bm{a}^{\rm T}\left\{ \bm{1}-\nu^2_r \bm{G}(\nu_r)\right\}
\left\{ \bm{1}-\nu^2_r \bm{G}(\nu_r)\right\} \bm{a} \nonumber \\
&+(i\nu_r)(-i\nu_r) \bm{b}^{\rm T}\bm{G}(\nu_r)\bm{G}(\nu_r)\bm{b}\Big],
\label{dxdtau2}
\end{align}
where we have used the fact that $\bm{G}(\nu_r)$ is even in $\nu_r$ and that the terms odd in $\nu_r$ vanish after the summation.
Using Eq.~(\ref{xrsol}), we have
\begin{align}
\frac 1 2\int_0^{\beta\hbar} d\tau \;  \bm{x}^{\rm T}(\tau)\bm{\Lambda}\bm{x}(\tau)=
\frac {1} {2\beta\hbar}\sum_{r=-\infty}^\infty \Big[ (-i\nu_r)(i\nu_r)\bm{a}^{\rm T} \bm{G}(\nu_r) \bm{\Lambda}\bm{G}(\nu_r)\bm{a}
+\bm{b}^{\rm T} \bm{G}(\nu_r)\bm{\Lambda}\bm{G}(\nu_r)\bm{b}\Big],
\end{align}
and from Eqs.~(\ref{influence}) and (\ref{La})
\begin{align}
\sum_{\alpha\in\mathcal{B}}\Psi_\alpha[x_\alpha]=& \frac M{2\beta\hbar}\sum_{r=-\infty}^\infty\sum_{\alpha\in\mathcal{B}}
 x_{\alpha,-r} \zeta^{(\alpha)}_r x_{\alpha,r} =\frac M{2\beta\hbar}\sum_{r=-\infty}^\infty
\bm{x}^{\rm T}_{-r} \bm{\Sigma}(\nu_r) \bm{x}_r \nonumber\\
=&\frac {M} {2\beta\hbar}\sum_{r=-\infty}^\infty \Big[ (-i\nu_r)(i\nu_r)\bm{a}^{\rm T} \bm{G}(\nu_r) \bm{\Sigma}(\nu_r)\bm{G}(\nu_r)\bm{a}
+\bm{b}^{\rm T} \bm{G}(\nu_r)\bm{\Sigma}(\nu_r)\bm{G}(\nu_r)\bm{b}\Big],
\end{align}
Adding these three contributions in Eq.~(\ref{mfg_path_mul}) and recalling the definition of $\bm{G}$ in Eq.~(\ref{green_mul}), we have
\begin{align}
&\langle \bm{\bar{x}}\vert \hat{\rho}_{\rm mG} \vert \bm{\bar{x}'}\rangle 
=C\exp\Big[ -\frac M{2\beta\hbar^2}\sum_{r=-\infty}^\infty   \left[ 
\bm{a}^{\rm T}\left\{ \bm{1}-\nu^2_r \bm{G}(\nu_r)\right\}\bm{a} +  \bm{b}^{\rm T} \bm{G}(\nu_r) \bm{b} \right] \Big],
\end{align}
Inserting Eqs.~(\ref{ajsol}) and (\ref{bjsol}) into this, we finally arrive at Eq.~(\ref{mfg_element_mul}).
\end{widetext}


\section{Diagonalization of Matrices for a Chain of Oscillators}
\label{app:4}

We first note that the matrices $\bm{\Lambda}$ and $\bm{V}$ in Eq.~(\ref{lambdav}) share the same eigenvectors. 
By direct application, we can check that
the eigenvectors of $\bm{V}$ in Eq.~(\ref{vmatrix}) are given up to normalization constant  in the form \cite{serafini2017quantum}
\begin{align}
\bm{t}_k\sim (1, e^{2\pi i  k/N}, e^{2\pi i (2k)/N},\cdots,e^{2\pi i (N-1)k/N})^{\rm T},
\end{align}
for $k=0,1,2,\cdots, N-1$ with eigenvalue $\lambda_k =2\cos (2\pi k/N)$.
The eigenvalue of $\bm{\Lambda}$ is then given by Eq.~(\ref{evalue}).
We note that there are two-fold degeneracies as $\lambda_{N-k}=\lambda_k$.
Below, we construct the orthonormal set of eigenvectors for the cases of even and odd $N$ separately. 

For $N$ odd, the space of $\lambda_0$ is one dimensional with normalized eigenvector
\begin{align}
\bm{t}_0=\frac{1}{\sqrt{N}}(1,1,\cdots,1)^{\rm T}.
\label{t0}
\end{align} 
On the other hand, the space of $\lambda_k=\lambda_{N-k}$ with $k=1,2,\cdots, (N-1)/2$
is doubly degenerate. We can form a linear combination of $\bm{t}_k$ and 
$\bm{t}_{N-k}$ which is proportional to $(1, e^{-2\pi i  k/N}, e^{-2\pi i (2k)/N},\cdots,e^{-2\pi i (N-1)k/N})^{\rm T}$ to make  a real eigenvector for 
this eigenvalue.
These are given by
\begin{align}
\bm{c}_k= \sqrt{\frac{2}{N}}\Big(1, \cos(\frac{2\pi}{N}k), & \cos(\frac{2\pi}{N}2k), \cdots\nonumber \\
&\cdots, \cos(\frac{2\pi}{N}(N-1)k)\Big)^{\rm T},
\label{ck} 
\end{align}
and
\begin{align}
\bm{s}_k= \sqrt{\frac{2}{N}}\Big(0, \sin(\frac{2\pi}{N}k), & \sin(\frac{2\pi}{N}2k),\cdots, \nonumber \\
&\cdots \sin(\frac{2\pi}{N}(N-1)k)\Big)^{\rm T}
\label{sk}
\end{align}
We can easily show that the collection of $\bm{t}_0, \{ \bm{c}_k\}$ and $\{ \bm{s}_k \}$ for $k=1,2,\cdots,(N-1)/2$ form an orthonormal set.
Now we relabel these vectors by using $\{ \bm{u}_k \}$ for $k=0,1,2,\cdots,N-1$ such that
$\bm{u}_0=\bm{t}_0$ and $\bm{u}_k=\bm{c}_{k}$ for $k=1,2,\cdots, (N-1)/2$. For the remaining part $k=(N+1)/2, (N+3)/2,\cdots,N-1$, 
we assign $\bm{u}_k=\bm{s}_{N-k}$. Then the set $\{ \bm{u}_k \}$ with $k=0,1,2,\cdots,N-1$ is the desired orthonormal set of eigenvectors
of $\bm{V}$ corresponding to eigenvalue $\lambda_k$.

Therefore, we can form an orthogonal matrix $\bm{R}$ by putting the eigenvectors $\{\bm{u}_k\}$ on the columns as
\begin{align}
\bm{R}=&\left( \bm{u}_0 \vert \bm{u}_1 \vert \cdots \vert \bm{u}_{N-1}\right) \nonumber \\
=&\left( \bm{t}_0 \vert \bm{c}_1 \vert \cdots \vert \bm{c}_{(N-1)/2}\vert \bm{s}_{(N-1)/2}\vert\cdots\vert \bm{s}_{1}\right) .
\end{align}
This matrix can then be used to diagonalize $\Lambda$ as
$\bm{R}^{\rm T}\bm{\Lambda}\bm{R}=\mathrm{diag}(M\Omega^2_0,\cdots,M\Omega^2_{N-1})$. 

Using this matrix $\bm{R}$, we can calculate the zeroth order Green's function in Eq.~(\ref{matrixR})  as
\begin{align}
G^{(0)}_{ij}(z)=&\frac{t^i_0 t^j_0}{z^2+\Omega^2_0}+\sum_{k=1}^{(N-1)/2} \frac{c^i_k c^j_k+s^i_k s^j_k}{z^2+\Omega^2_k}  \\
=& \frac 1 N \frac{1}{z^2+\Omega^2_0}+\frac 2 N \sum_{k=1}^{(N-1)/2} \frac{\cos(2\pi (i-j)k/N)}{z^2+\Omega^2_k} \nonumber .
\end{align}
This can be easily seen to be equivalent to Eq.~(\ref{G0}).

The case of even $N$ is similar. We first note that the spaces corresponding to $\lambda_0$ and $\lambda_{N/2}$ are both one dimensional
with eigenvectors $\bm{t}_0$ in Eq.~(\ref{t0}) and
\begin{align}
\bm{t}_{N/2}=\frac{1}{\sqrt{N}}(1,-1,1,-1,\cdots,1,-1)^{\rm T},
\end{align} 
respectively. For the doubly degenerate spaces
corresponding to $k=1,2,\cdots,N/2-1$, the eigenvectors are again given by
$\bm{c}_k$ and $\bm{s}_k$ defined in Eqs.~(\ref{ck}) and (\ref{sk}). Therefore, the orthonormal set $\{\bm{u}_k\}$ with
$k=0,1,\cdots, N-1$ is given by
$\bm{u}_0=\bm{t}_0$, $\bm{u}_{N/2}=\bm{t}_{N/2}$, and $\bm{u}_k=\bm{c}_{k}$ for $k=1,2,\cdots, N/2-1$.
For $k=N/2+1, N/2+2,\cdots,N-1$, we take $\bm{u}_k=\bm{s}_{N-k}$.
In this case, the orthogonal matrix $\bm{R}$ we are after takes the form
\begin{align}
\bm{R}
=\left( \bm{t}_0 \vert \bm{c}_1 \vert \cdots \vert \bm{c}_{N/2-1}\vert \bm{t}_{N/2}\vert \bm{s}_{N/2-1}\vert\cdots\vert \bm{s}_{1}\right) .
\end{align}
We therefore have in this case
\begin{align}
G^{(0)}_{ij}(z)
&= \frac 1 N \frac{1}{z^2+\Omega^2_0}+\frac 2 N \sum_{k=1}^{N/2-1} \frac{\cos(2\pi (i-j)k/N)}{z^2+\Omega^2_k} \nonumber  \\
&+\frac 1 N  \frac{(-1)^{i+j}}{z^2+\Omega^2_{N/2}}.
\end{align}
Again this can equivalently be written as Eq.~(\ref{G0}). 

\section{Evaluation of the Density Matrix in the Ultra Strong Coupling Limit}
\label{app:5}

We closely follow the procedure described in Appendices \ref{app:2} and \ref{app:3} to evaluate the matrix 
element of $\rho_{\rm USC})$ given in Eq.~(\ref{rho_USC}).
The stationary path for the path integral satisfies 
\begin{align}
-M\frac{d^2 x_1(\tau)}{d\tau^2}+M\Omega^2 x_1(\tau)-2M\lambda \bar{x}_0=0.
\end{align}
This equation is solved 
in terms of the Fourier component defined in Eq.~(\ref{xexpand_mul}) as
\begin{align}
x_{1,r}=g(\nu_r)\left[ 2\lambda\bar{x}_0 (\beta\hbar)\delta_{r,0} -i\nu_r a -b \right].
\label{xrsol_usc}
\end{align}
where 
\begin{align}
g(\nu_r)=\frac 1 {\nu^2_r+\Omega^2}.
\end{align}
Here $a$ and $b$ result from the discontinuities at the boundaries at $\tau=r \beta \hbar$ and will be determined
from the specific boundary conditions we require as we have done in Appendix \ref{app:2} (see Eq.~(\ref{xr_sol_app})).
The discontinuity of $x_1(\tau)$ at $\tau=0$ is responsible for $a$, which is given by
\begin{align}
a=\bar{x}_1^\prime-\bar{x}_1.
\end{align}
Now following the same argument leading up to Eqs.~(\ref{eq_b1}) and (\ref{eq_b2}), we have
\begin{align}
&\bar{x}_1^\prime=\frac a 2  + \frac 1 {\beta\hbar}\sum_{r=-\infty}^\infty g(\nu_r)\left[ 2\lambda \bar{x}_0(\beta\hbar)\delta_{r,0}-b\right],\\
&\bar{x}_1=-\frac a 2  + \frac 1 {\beta\hbar}\sum_{r=-\infty}^\infty g(\nu_r)\left[ 2\lambda \bar{x}_0(\beta\hbar)\delta_{r,0}-b\right].
\end{align}
Adding these two, we have
\begin{align}
b=\frac{1}{F}\left[ \frac{2\lambda\bar{x}_0}{\Omega^2}-\frac{\bar{x}_1+\bar{x}_1^\prime}{2}\right],
\end{align}
where
\begin{align}
F\equiv\frac 1 {\beta\hbar}\sum_{r=-\infty}^\infty g(\nu_r)=\frac 1 {2\Omega}\coth \left( \frac{\beta\hbar\Omega}{2} \right).
\end{align}

We now put the stationary path solution Eq.~(\ref{xrsol_usc}) in the action in Eq.~(\ref{S_USC}). Following the same steps leading to 
Eqs.~(\ref{dxdtau}) and (\ref{dxdtau2}), we have
\begin{align}
&\frac{M}{2}\int_0^{\beta\hbar}  d\tau\; \left( \frac{d x_1(\tau)}{d\tau}\right)^2 \\
=& \frac M 2 \frac 1 {\beta\hbar} \sum_{r=-\infty}^\infty
\Big[ (1-\nu^2_r g(\nu_r))^2 a^2 +\nu^2_r g^2(\nu_r) b^2\Big]  . \nonumber
\end{align}
We also obtain
\begin{align}
&\frac {M\Omega^2} 2 \int_0^{\beta\hbar}  d\tau\; x^2_1(\tau) \\
=&\frac {M\Omega^2} 2 \Bigg[ \frac 1 {\beta\hbar} \sum_{r=-\infty}^\infty
g^2(\nu_r)(\nu^2_r a^2 +b^2) 
-\frac{4\lambda \bar{x}_0}{\Omega^4}b + \frac{4\lambda^2 \beta\hbar \bar{x}^2_0}{\Omega^4}\Bigg], \nonumber
\end{align}
and
\begin{align}
-2M\lambda\bar{x}_0  \int_0^{\beta\hbar}  d\tau\; x_1(\tau) 
=- \frac{2M\lambda\bar{x}_0}{\Omega^2} [2\lambda \bar{x}_0\beta\hbar- b].
\end{align}
Adding these three terms together with $(M/2)\Omega^2 \bar{x}^2_0$ term, we have the stationary path expression for
the action in Eq.~(\ref{S_USC}) as
\begin{align}
S_{\rm USC}&=\frac M 2 [A a^2 +F b^2 +\beta\hbar\Omega^2\{1-\frac{4\lambda^2}{\Omega^4}\}\bar{x}^2_0] \nonumber \\
&=\frac M 2 \Big[A (\bar{x}_1^\prime-\bar{x}_1)^2 +\frac 1 F \left(\frac{\bar{x}_1+\bar{x}_1^\prime}{2}- \frac{2\lambda}{\Omega^2}\bar{x}_0\right)^2 \nonumber \\
& ~~~~~~~~ +\beta\hbar\Omega^2\left(1-\frac{4\lambda^2}{\Omega^4}\right)\bar{x}^2_0\Big] 
\end{align}
where 
\begin{align}
A\equiv \frac 1 {\beta\hbar}\sum_{r=-\infty}^\infty (1-\nu^2_r g(\nu_r))=\frac {\Omega} {2}\coth \left( \frac{\beta\hbar\Omega}{2} \right).
\end{align}
Putting this expression for the action into Eq.~(\ref{rho_USC}), we obtain Eq.~(\ref{matrix_usc}).

\section{The large-$\gamma$ limit of $\langle \hat{p}^2_0\rangle_{\rm mfG}$}
\label{app:6}

From Eq.~(\ref{pp_full}), we have
\begin{align}
\langle \hat{p}^2_0\rangle_{\rm mfG}=\frac  {M}{\beta}[1-2\sum_{r=1}^\infty  \{1-\nu^2_r G_{00}(\nu_r)\}].
\label{app6eq}
\end{align}
If we use the Ohmic bath, we have from Eqs.~(\ref{full_green}) and (\ref{N2G00})
\begin{align}
G_{00}(z) 
=\frac{(z+\omega_D)(z^2+\Omega^2)}{P(z) },
\end{align}
where
\begin{align}
P(z)=(z+\omega_D)(z^2+\Omega^2_0)(z^2+\Omega^2_1)+2\gamma\omega_D z(z^2+\Omega^2)
\end{align}
for nonnegative frequency $z=\nu_r\ge 0$. 
In order to evaluate $\langle \hat{p}^2_0\rangle_{\rm mfG}$ from Eq.~(\ref{app6eq}), we need to decompose
\begin{align}
1-z^2 G_{00}(z)=\sum_{i=1}^5 \frac{D_i}{z+z_i},
\end{align}
where $-z_i$ is the $i$-th root of $P(z)=\prod_{i=1}^5 (z+z_i)$ and the coefficient $D_i$ is determined from
the method of residues. The large-$\gamma$ behaviors of $z_i$ are given by
$z_1\sim O(\gamma^{-1})$, $ z_2\sim -i\Omega+O(\gamma^{-1})$, $ z_3\sim i\Omega+O(\gamma^{-1})$, and
\begin{align}
&z_4\sim -i\sqrt{2\omega_D\gamma }+O(\gamma^{-1/2}) \\
&z_5\sim i\sqrt{2\omega_D\gamma}+O(\gamma^{-1/2}) .
\end{align} 
We can easily establish that in the large-$\gamma$ limit, the leading order behavior of $D_1, D_2$ and $D_3$ are of $O(1)$ and
\begin{align}
&D_4\sim -i\sqrt{\frac{\omega_D\gamma}{2}} ,~~ D_5\sim i\sqrt{\frac{\omega_D\gamma}{2}} .
\end{align}
Since $D_4+D_5=0$ in this limit,
 the sum over the positive frequency in Eq.~(\ref{app6eq}) can be written in terms of the digamma function $\psi(x)$ \cite{arfken}.
 If we write $\nu_r=r\nu$ with $\nu=2\pi/(\beta\hbar)$, we have
\begin{align}
\langle p^2_0\rangle_{\rm mfG}&\sim \frac{M}{\beta}\left[-\frac 2{\nu}\sum_{i=4}^5 D_i \psi(1+\frac{z_i}{\nu})\right] \nonumber \\
& \sim -\frac{M\hbar}{\pi} \sum_{i=4}^5 D_i \ln z_i \nonumber\\
&\sim M\hbar\sqrt{\frac{\omega_D\gamma}{2}},
\end{align}  
where we have used the asymptotic behavior of the digamma function \cite{arfken}.


%


\end{document}